\documentclass[useAMS]{mn2e}

\input psfig.sty
\def\etal{et al.}

\title[Environmental effects on galaxies and AGN]{{  Local and Global }Environmental effects on galaxies and AGN}
\author[Padilla, Lambas \& Gonz\'alez]{Nelson Padilla$^{1}$\thanks{E-mail:
npadilla@astro.puc.cl}, Diego Garc\'\i a Lambas$^{2}$, Roberto Gonz\'alez$^{1}$\\
$^{1}$Departamento de Astronom\'\i a y Astrof\'\i sica, Pontificia Universidad Cat\'olica de Chile, Santiago, Chile.\\
$^{2}$Instituto de Astronom\'\i a Te\'orica y Experimental,Conicet-UNC, C\'ordoba, Argentina.}
\begin{document}

\date{Accepted ????. Received ????; in original form ????}

\pagerange{\pageref{firstpage}--\pageref{lastpage}} \pubyear{2008}

\maketitle

\label{firstpage}

\begin{abstract}
We study the properties of SDSS galaxies with and without AGN detection as a function of the local and global environment measured via the local density, the mass of the galaxy host group (parameterised by the group luminosity) and distance to massive clusters.  {  Our results can be divided in two main subjects, the environments of galaxies and their relation to the assembly of their host haloes, and the environments of AGN.  (i)} For the full SDSS sample, we find indications that the local galaxy density is the most efficient parameter to separate galaxy populations, but we also find that galaxies at fixed local density show some remaining variation of their properties as a function of the distance to the nearest cluster of galaxies (in a range of $0$ to $10$ cluster virial radii).  These differences seem to become less significant if the galaxy samples are additionally constrained to be hosted by groups of similar total luminosity.   {  If instead of fixing the local density, the mass of the host group is held fixed at a given value, the fraction of red galaxies also increases as the distance to clusters diminishes, indicating that neither the local density or the host halo mass contain all the information on the environment}.  (ii) In AGN host galaxies, the morphology-density relation is much less noticeable when compared to the behaviour of the full SDSS sample, indicating a lack of sensitivity to the host group mass during the AGN phase probably due to the higher typical luminosities of the AGN hosts.  In order to interpret this result we analyse {  control samples constructed using galaxies with no detected AGN activity} with matching distributions of redshifts, {  stellar masses}, r-band luminosities, $g-r$ colours, concentrations, local densities, host group luminosities, {  and fractions of central and satellite galaxies; the aim in using the control sample is to detect any correlations between the AGN detection and other AGN host properties that are unrelated to the AGN selection}.  The control samples also show a similar small dependence on the local density indicating an influence from the AGN selection, but their colours are {  slightly bluer} compared to the AGN hosts regardless of local density.  {  Furthermore}, {\it even when the local density is held fixed} at intermediate or high values, and the distance to the closest cluster of galaxies is allowed to vary, AGN control galaxies away from clusters tend to be bluer than the AGN hosts.  {  However}, AGN in bright, low concentration hosts (i.e.  disky morphologies) are bluer than galaxies in the control sample, connecting the presence of discs to AGN activity {  even under a controlled comparison between active and inactive galaxies.}
\end{abstract}


\section{Introduction}

The description of the large-scale structure in the Universe, and how it relates
to the properties of galaxies in different environments has become more intricate in
the last few years.  {  In order to understand the environmental dependence of galaxy
properties it is now necessary to combine results where the environment comes from
local density estimates, the host halo mass, the distance to cluster centres, together
with a theoretical understanding of what could influence the SF activity and morphological
transformations in galaxies.  This introduction will first review the assembly effects
on the galaxy population of haloes of equal mass, then the results from the literature on 
the environment studies of galaxies, and finally the properties of the hosts of AGN, and
their dependence on environment.}

\subsection{Properties of galaxies in equal mass haloes: assembly effects}

Up to only a few years back it was believed that the mass of a dark-matter halo was the only
important parameter that defined its clustering properties (see for instance
Padilla \& Baugh 2002; Sheth, Mo \& Tormen 2001; Mo \& White 1996) and galaxy population (e.g. Cole
et al., 2000).
The latter has been studied in the framework of the Halo Model (Cooray \& Sheth 2002; 
Cooray 2005, 2006, 
Jing, Mo \& B\"orner, 1998, Peacock \& Smith, 2000),
whereby the typical number of galaxies per dark-matter halo,
typical central and satellite luminosities, {  star formation activity}, and possibly their colours 
depend solely on the host halo mass
({  Weinmann et al. 2006}; Wang et al. 2008).  
However, recent studies on the clustering of dark-matter haloes found that the age and assembly history
of haloes of similar mass also play a role in their clustering amplitude.  The dependence on halo age
was first reported by Gao, Springel \& White (2007), and has been studied in more detail 
finding a general dependency on the way in which a halo is assembled by Gao \& White (2007),
Jing, Suto \& Mo (2007), Croton, Gao \& White (2007), Wechsler et al., (2007), and Li, Mo \& Gao (2008).
The assembly bias has also been detected in observations by Wang et al. (2008); their studies
only concentrate on the clustering amplitude of different samples selected according to
group colour which they associate with group age.
In addition to a difference in clustering amplitude, Zapata et al. (2009) demonstrated that groups
of similar mass and different assembly histories also show important differences in their
population of galaxies, an aspect that can be added to current halo model and conditional luminosity
functions, and that can also be used to improve the modeling and understanding of the evolutionary
processes that shape galaxies in the observed Universe.

{ Extending this particular problem,
Wang et al. (2009) found a sub-population of red, dwarf isolated
galaxies that are not satellites of larger haloes but that are concentrated around
their nearest massive halo.   Such isolated dwarfs are expected to be able to form
stars and therefore show blue colours.  These red, dwarf galaxies would fit within 
the galaxy formation scenario if they inhabit subhaloes that have been
expelled from clusters and have acquired their red colours earlier, when they were under
the effects that satellites suffer in Clusters such as strangulation, ram-pressure
and tidal stripping, processes that quench the star formation activity.  
On the theoretical side, Wang, Mo \& 
Jing (2009) find that some small fraction of the assembly bias effect
comes from such subhaloes.}

These results however, concern only the population of galaxies inside individual
dark-matter haloes (or ejected sub-haloes).  

\subsection{Measurements of environment and the properties of galaxies}

On the
large-scale structure side, starting with the pioneering work of Dressler et al. (1980),
the location of a galaxy has been shown to have an influence on its average properties.  Using
large redshift surveys, Balogh et al. (2004) found a smooth transition
from high density regions populated by red galaxies, towards blue dominated low density regions.
These results pointed to the density of the region where galaxies live as being responsible
for important characteristics of the galaxy population.  This was also confirmed
at higher redshifts in the Sloan Digital Sky Survey (SDSS, 
York et al. 2000), by O'Mill, Padilla \& Lambas (2008).  
Numerical simulations following the evolution of dark-matter, with galaxy populations
imprinted onto them via semi-analytic models, have shown that the prevalence of red galaxies
in high density regions can be naturally achieved in a $\Lambda$CDM hierarchical cosmology
(Croton et al. 2006; Bower et al. 2006; Cattaneo et al. 2006, 2008; Lagos, Cora \& Padilla 
2008; Lagos, Padilla \& Cora 2009).
A remaining question with respect to this general behaviour resided in whether this population
change was due to local or global effects,{  that is, whether the dependence of properties
on the local density responds mainly to the dependence on halo mass alone, or also to assembly
and other effects such as ejection from hosts}.  

Kauffmann et al. (2004) found little evidence of a dependence of star formation (SF) indicators
on the environment measured over scales larger than $1$h$^{-1}Mpc$.  However, more
recent results around voids in the SDSS by Ceccarelli, Padilla \& Lambas (2008) suggested that
the local density is not the only parameter defining a galaxy population.  They showed that
at fixed local density, 
the fraction of blue galaxies increases towards the regions corresponding to void walls;
this represented a first measurement of a large-scale modulation of star formation in galaxies.
This behaviour was later confirmed in numerical simulations by Gonz\'alez \& Padilla (2009),
who analysed semi-analytic galaxies from the SAG model (Lagos, Cora \& Padilla 2008; Lagos, Padilla \& Cora 2009),
and found that this was also the case for galaxies of fixed local density, at increasingly
larger distances from clusters of galaxies, or closer to voids.  Furthermore, they found
that the reason behind this change at fixed local density, is mostly due to the effect
of the mass of the host halo, with a remnant variation that
can be tied to the assembly history of the haloes hosting these galaxies.  
The reason for a dependence on mass is related to the different physical
processes taking place in haloes of different masses.  For instance,
at very low halo masses, $M<10^{10}$h$^{-1}M_{\odot}$, reionization from a UV radiation field plays
an important role in the photo-ionization of baryons; this effect along with stellar feedback
reduces significantly the baryon fraction available to form stars in such low mass haloes (Tassis et al. 2002).
The reason for a dependence on assembly is that, for instance, more recent merger activity can have an important
impact on the colours of galaxies.
Therefore, even though local processes may extend out to $\sim 1$h$^{-1}$Mpc, global processes
affecting the underlying mass function and the rate of growth of haloes (assembly)
help extend the variation of galaxy properties to
much larger scales, up to tens of Mpc.

\subsection{AGN host properties, and their dependence on environment}

Regarding the environmental dependence of galaxies with active galactic nuclei (AGN) there
is still a debate on the  frequency of occurrence of AGN as a function of the local density.  For
instance, Miller et al. (2003) found that the fraction of galaxies with AGN appeared to be constant
with local density, even when dividing the sample according to the host morphology into
spiral and elliptical galaxies (see also Dressler et al. 1999).
Using the SDSS, Kauffmann et al. (2004) confirmed this result only for AGN with weak emission in
the OIII line.  They claim that for high OIII emitters the fraction of 
galaxies with AGN decreases as the local density increases (see also Popesso \& Biviano
2006).  
It should be noted that these studies do not reach densities as high as
those in clusters of galaxies.  
Pasquali et al. (2009) studied the variation of the SF
and AGN activity of central and satellite galaxies as a function of their host
dark-matter halo mass (ranging from low mass groups to clusters of galaxies), 
to find a smooth and continuous transition from SF to AGN and to
radio activity as the halo mass { and the stellar mass} increase, for both centrals and 
satellites, the latter
showing lower activity than the former at all halo masses; { furthermore they show
that the dependence on environment (as indicated by the halo mass) is 
weaker than on the stellar mass of the host galaxy}.
Other studies of AGNs in clusters
show a higher frequency of X-ray sources than in the field
(Cappi et al. 2001; Molnar et al. 2002; Martini et al. 2002).
Furthermore, using clustering analyses, Gilli et al. (2003) also found higher correlation
amplitudes for
X-ray selected AGNs.  
Additionally, Martini, Sivakoff \& Mulchaey (2009) found that the fraction of AGN in clusters
of galaxies decreases for lower redshifts; this could help obtain a smooth transition between
the results from these X-ray detected AGN and those by Kauffmann et al. (2004).

There are also studies of the environment in which AGN are embedded, { which focus on 
the morphology and other characteristics of their hosts and their neighbors (including
their local density, colours, etc)}.
Choi et al. (2009) studied AGN hosts in the SDSS Data Release 5 
(DR5, Adelman-McCarthy et al. 2008).  They find that
late morphological types are the dominant hosts regardless of AGN power, and that
bluer late types host the most powerful AGN (also in agreement with Coldwell et al. 2009),
{  confirming earlier results by Kauffmann et al. (2003a), who additionally show that the
young stars in the powerful AGN reside preferentially away from the nucleus, and that their hosts have
suffered a starburst in the recent past.  Additionally, Kauffmann et al. (2003a) show that these
results do not depend on the Seyfert type (for equal OIII luminosity).  Going into more detail 
on the different AGN types, Kewley et al. (2006) show that the hosts of LINERs (low-ionization narrow
emission-line regions, characterised by low luminosities with respect to the Eddington value) 
are older, more massive, less dusty, less concentrated and have higher velocity dispersions
then Seyfert galaxies (characterised by higher $L/L_{EDD}$).}
Waskett et al. (2005) found that the
environments of AGN at $0.4<z<0.6$ are similar to those of normal galaxies at these same
redshifts, and consist of group-like environments.
Gilmour et al. (2007) confirm this result finding that
AGN in bright galaxies at $z=0.17$ (in a region corresponding to a supercluster) show a preference
for intermediate environments, and are preferentially surrounded by blue galaxies.
More recently in zCOSMOS (Lilly et al. 2007), Silverman et al. (2009) 
confirm that AGN tend to prefer intermediate environments such
as groups, and also find AGN hosted by massive 
galaxies with stellar masses $M_*>2.5\times 10^{10}M_{\odot}$ that
occur preferentially in low density environments where disruptive 
processes (mergers) are lessened.
Results on the environment of high redshift AGN sources 
(detected via X-ray counterparts of ACS images) are presented by 
Martel et al. (2007), who find that $z\simeq1$ AGN prefer
regions which are more crowded than for the typical galaxy in the field, giving clues on the 
possible nature of the hosts of these AGN; however, they also
warn about possible systematic effects in their results.
More recently, von der Linden et al. (2009, see also Montero-Dorta et al. 2009) 
study the variation of the fraction
of AGN hosts as a function of the distance to cluster centres in the SDSS.  They
analyse the time-scale over which the AGN activity shuts down and find that this coincides
roughly with the crossing time-scale of the clusters, and that this is similar to the
timescale of general SFR shut-down; {  note that their analysis does not discriminate the
changing local density as the distance to the cluster centres varies}.

Centering mostly on low accretion rate AGNs, 
these studies have helped shed light on the feeding and growing mechanisms of AGN, and
on their influence on the star formation (SF) activity of the host galaxies.
For instance,
Koulouridis et al. (2006) studied the origin of the AGN activity in relation to their immediate
environment.
Their analysis of bright IRAS galaxies in comparison to that of AGN helped them conclude that
close interactions drive fuel to the AGN, which produces obscured AGN activity.  They
infer that only when the close neighbor moves away, the AGN becomes unobscured.  Martini
et al. (2004) presents an extended review on the relation between AGN fueling and environment.
These results are in good agreement with morphological studies by 
Pasquali, van den Bosch \& Rix (2007)
who find that the AGN activity is associated with an increase in 
the fraction of galaxies with stellar discs, and Hao et al. (2009) who detect
an increase in the fraction of barred hosts with the presence of AGN.  Along the same
lines, Donoso et al. (2009) find, via clustering studies of radio-loud AGN, that there is
an influence from the gaseous content on scales of the host dark-matter halo on the
probability that an AGN is radio-loud and on the jet luminosity (see
Donoso,
Best \& Kauffmann 2009, for a study on the evolution of radio AGN with redshift).

\subsection{The object of this work}

In this paper we will focus on the variation of galaxy properties at different global
environments as traced simultaneously 
by the local density and by the distance to the nearest clusters of galaxies, {  which
as mentioned above should include both a dependence on the host dark-matter halo mass, halo assembly,
and other effects such as, for instance, dynamical coherence or ejection from larger haloes}.  
We will study galaxy colours, concentrations,
and colour-magnitude relations, for normal galaxies on the one hand and for galaxies with AGN on the other,
with the aim to provide further
clues on the interplay between large-scale structure, assembly history, and galaxy evolution.
{  The reader should bear in mind that the environmental behaviours of normal galaxies
and of AGN hosts will not be compared to one another due to the different selection effects
in play, but will be used to try to improve our understanding of the galaxy formation process.}
{  In the case of AGN hosts,} we {  want to stress the fact 
that we} will focus our analysis of the environmental dependence 
via variations with the environment instead of studying the fraction of galaxies
hosting AGNs.  In order to overcome the possible influence from selection effects associated
with the detection of AGN signatures in the spectra of the galaxies we will rely on the
definition of control samples that match several properties of the AGN hosts to mimic 
their selection function as best as possible.
{  Finally, 
we will use the galaxy colour as an indicator of star formation activity, bearing in mind
that this quantity also depends on star-formation history, metallicity and reddening.}

This paper is organised as follows, Section 2 presents the sets of observational data and 
definition of samples used in this
work; in Section 3 we explore
distributions of galaxy colours and concentrations for the full sample of galaxies and for AGN.
We present our results and compare them to the literature in Section 4, which also contains our conclusions.

\section{Data}

\subsection{SDSS Galaxies and AGN}
\label{sec:data}

We study galaxies from the spectroscopic SDSS, Data Release 7
(DR7, Abazajian \etal 2009),  which consists of $\sim 686,000$ galaxies
with measured spectra and photometry in five photometric bands, $ugriz$.  
In our analysis we restrict the full sample of SDSS galaxies to $z<0.1$, which leaves
a total of $307,254$ galaxies.  This redshift limit reduces the effects from
flux incompleteness; we correct for the remaining incompleteness effects
by weighting each galaxy by the inverse
of the maximum volume out to which it can be detected.

Several
galaxy parameters are provided by the automated SDSS pipeline out of which we 
make use of the Petrosian magnitudes and the
radii containing $50$ and $90$ percent of the total flux of each galaxy,
calculated in the $r$ photometric band.  We define the bulge concentration parameter
\begin{equation}
c_b=r_{90}^r/r_{50}^r,
\end{equation}
which has been proposed as a morphological discriminator such that low (high)
concentration galaxies consist preferentially of spiral (elliptical) types
(Shimasaku et al. 2001; Strateva et al. 2001).

We use the OIII, NII, H$\beta$ and H$\alpha$ line ratios measured by Kauffmann et 
al. (2003b) initially for the SDSS Data Release 1 (DR1, Abazajian et al. 2003) updated
to the Data Release 4 (DR4), to select AGN in our sample, to compare their characteristics
with that of the full galaxy population.  We follow Kauffmann et al. (2003b) and define
a galaxy as an AGN if,
\begin{equation}
\log([OIII]/H\beta)>0.61/(\log([NII]/H\alpha)-0.05)+1.13,
\end{equation}
{  with all four lines satisfying $S/N>3$.  There are $36107$ AGN with $z<0.1$ in this sample.}
According to Kauffmann et al. (2003b), none of the AGN emission features affect the
r-band photometry or the concentration parameter of the galaxies in our samples.
Our analysis of AGN hosts will also take into account the redshift dependent detection limit
by weighting each galaxy by the maximum volume out to which it would be detected given the magnitude
limit of the survey $r=17.77$.  Hao et al., (2005) note that this limit would only be slightly
different for the detection of the AGN spectral lines, estimating corrections of less than a $10\%$.

In the remainder of this paper we will consider galaxies with no (or undetected) 
AGN activity as those with null luminosity in the OIII emission line in the spectroscopic DR4 sample
($130828$ galaxies out to $z=0.1$).

\subsection{SDSS Group catalogue}
We use the group catalogue constructed from the 
SDSS-DR6 by Zapata et al. (2009), updated to the full DR7.  This catalogue is
based on a friends-of-friends algorithm which uses a
varying projected linking length $\sigma$, with $\sigma_0=0.239$ h$^{-1}$Mpc 
and fixed radial linking length 
$\Delta V=450 \ {\rm kms^{-1}}$.  These parameters correspond to the values found 
by Merch\'an \& Zandivarez (2005) to
produce a reasonably complete sample ($95\%$) with low contamination ($\la 8\%$).
The quality of the group catalogue has been analysed by applying the same group
identification algorithm to a SDSS mock catalogue (Zapata et al. 2009).
We acknowledge previous group compilations using the FOF technique on the 
SDSS DR3 (Merch\'an \& Zandivarez 2005) and DR4 (Yang et al. 2007), and using
percolation algorithms (Berlind et al. 2006) on the NYU-VAGC (Blanton et al. 2005). 
We adopt the Zapata et al. groups since this choice allows us
to control the measurements of group centre, mass {  (or group luminosity)}
and membership.  In addition, we are also able to use a
larger sample of galaxies, that of the SDSS DR7.  
The sample of groups extends out to $z=0.12$, with a total of $15,140$ groups
with at least four galaxy members. 

\begin{figure*}
\begin{picture}(340,255)
\put(-80,125){\psfig{file=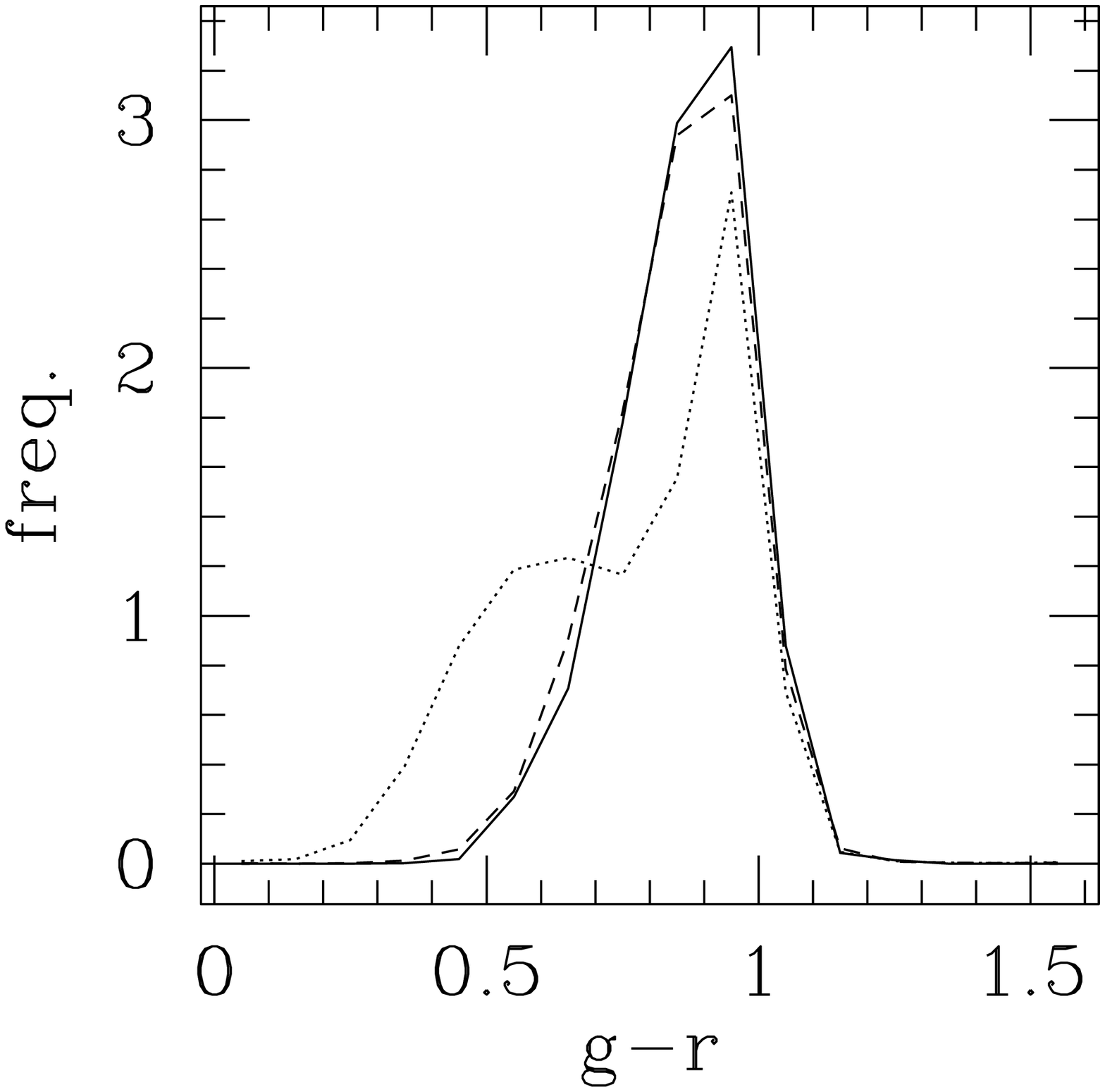,width=4.5cm}}
\put(-80,-10){\psfig{file=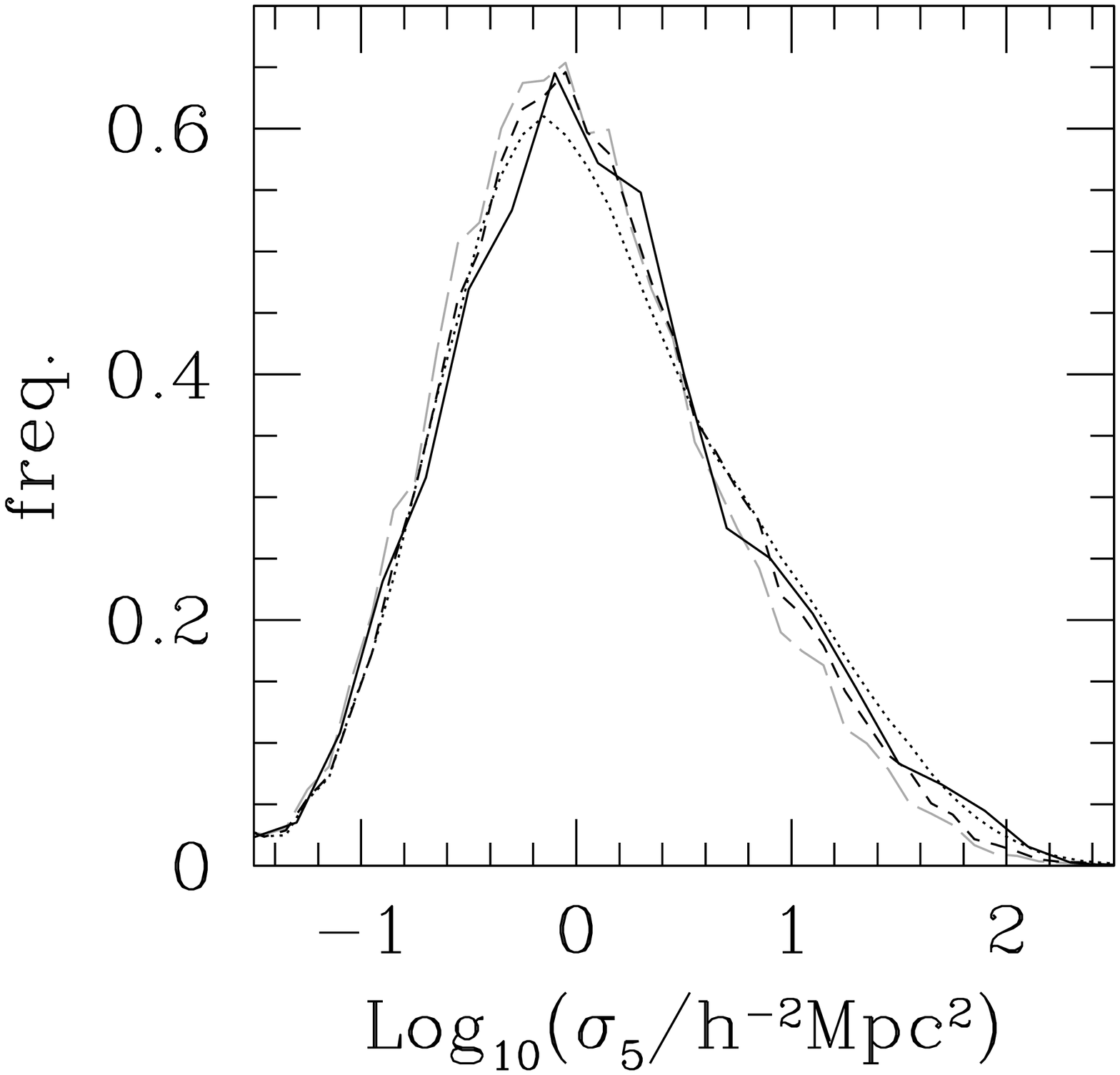,width=4.5cm}}
\put(45,125){\psfig{file=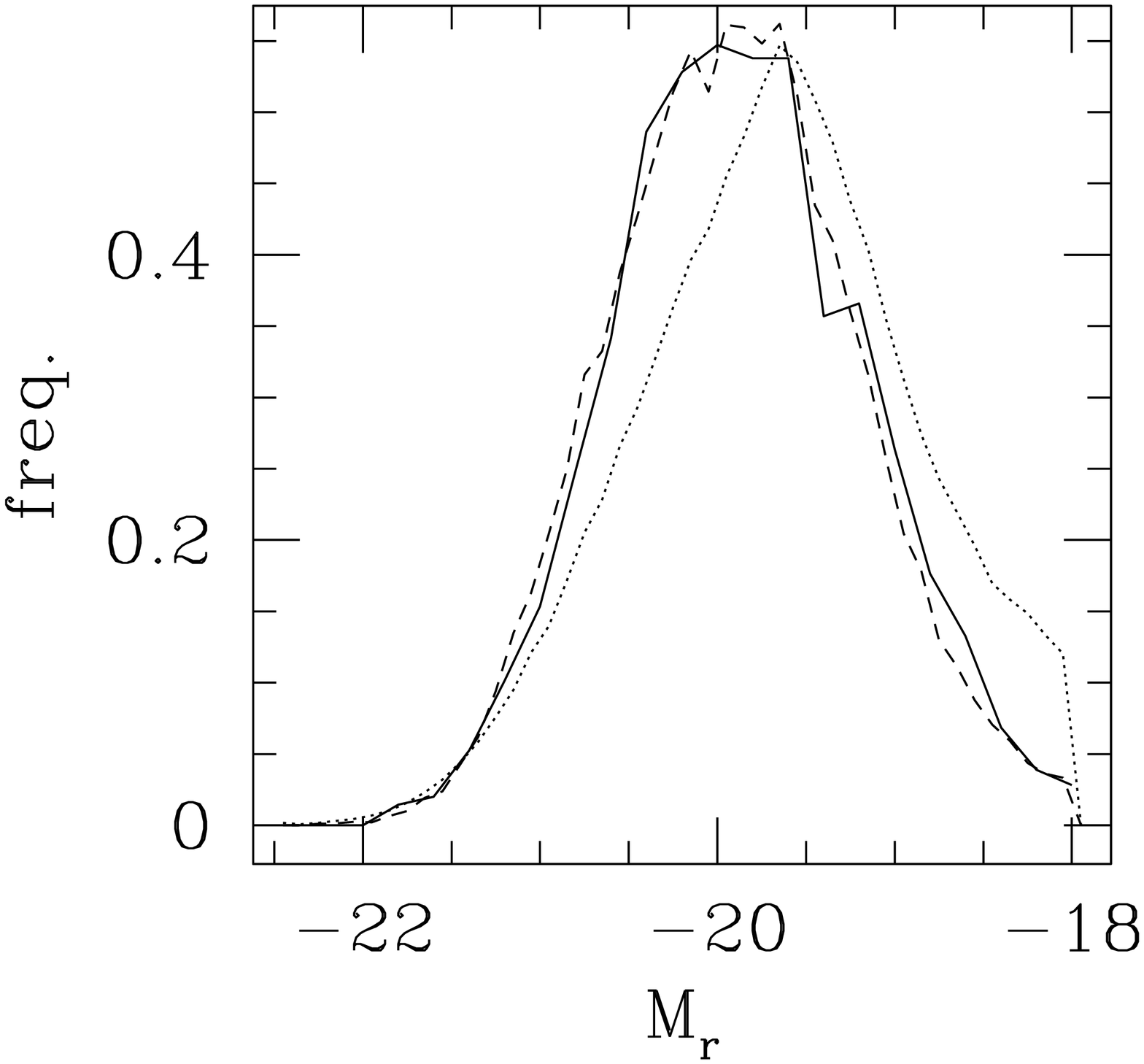,width=4.5cm}}
\put(45,-10){\psfig{file=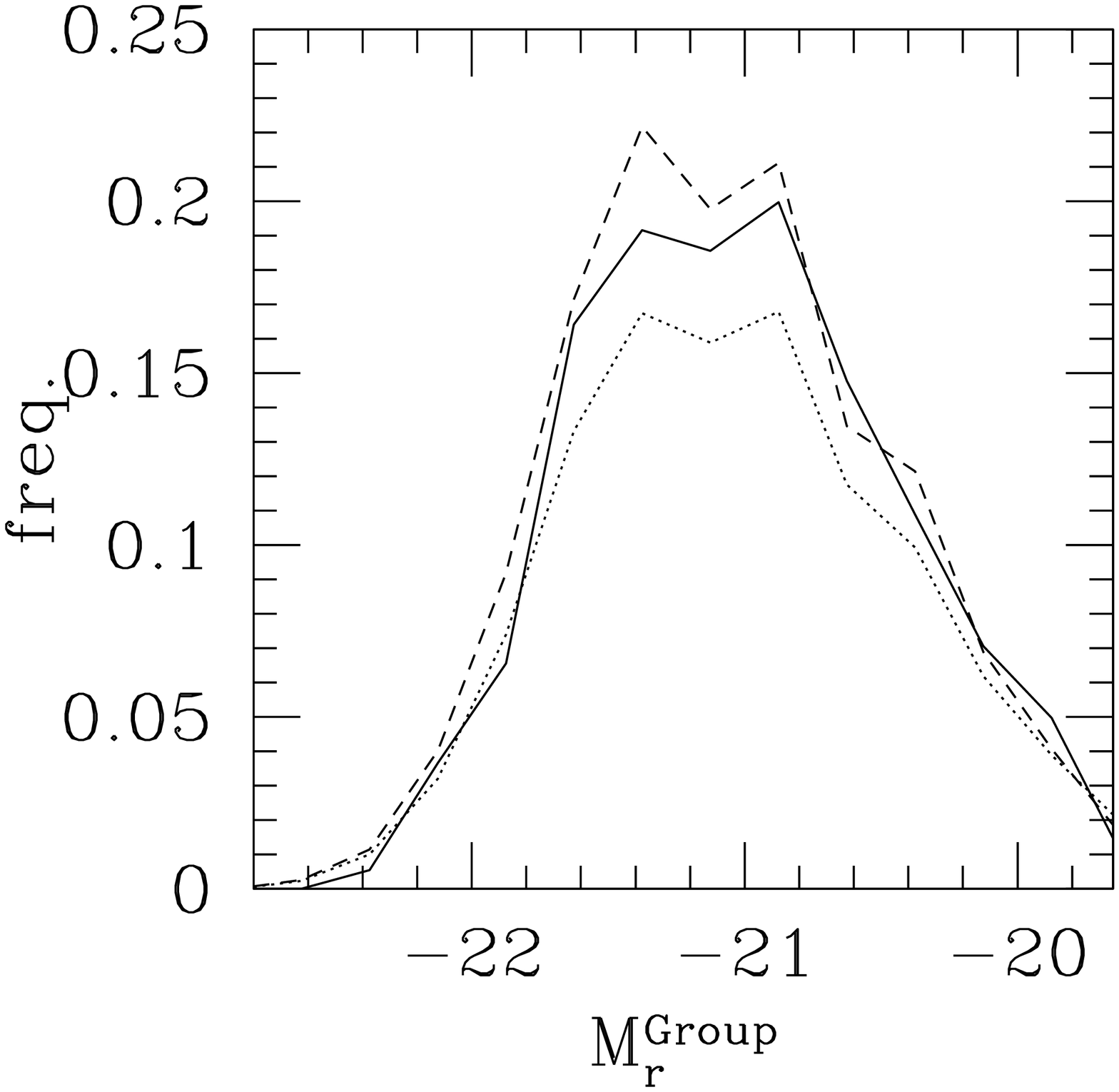,width=4.5cm}}
\put(170,125){\psfig{file=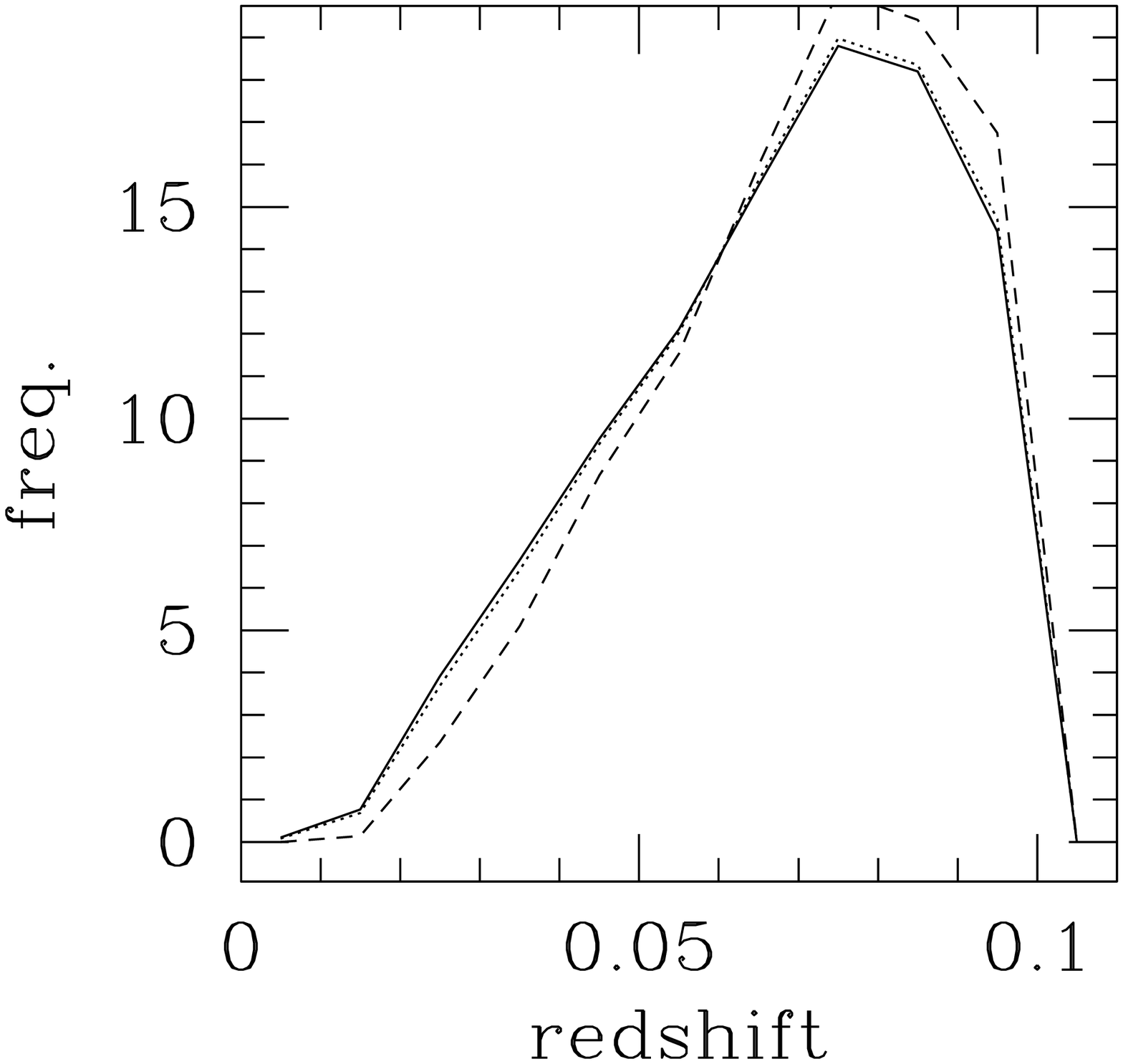,width=4.5cm}}
\put(170,-10){\psfig{file=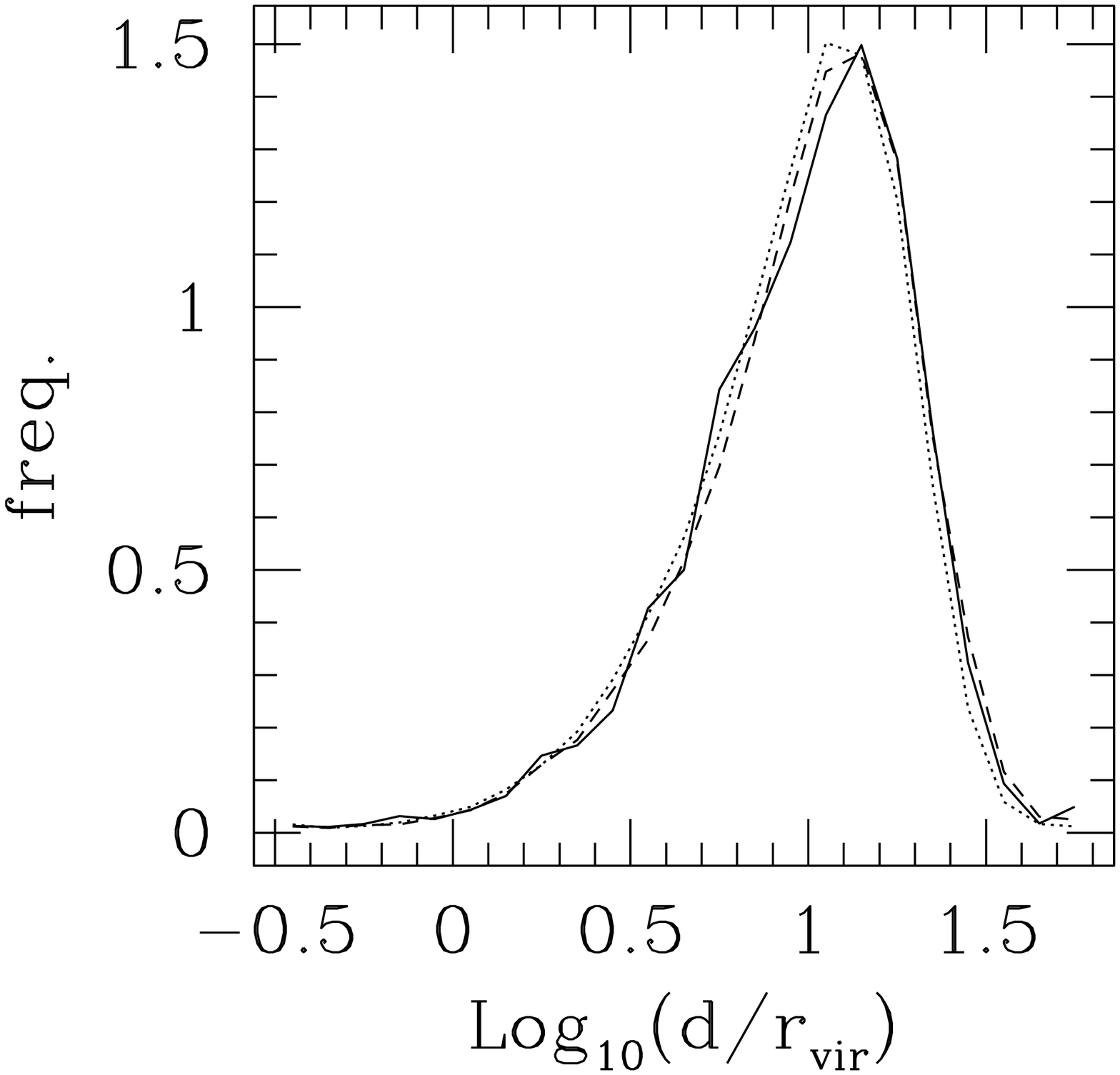,width=4.5cm}}
\put(295,125){\psfig{file=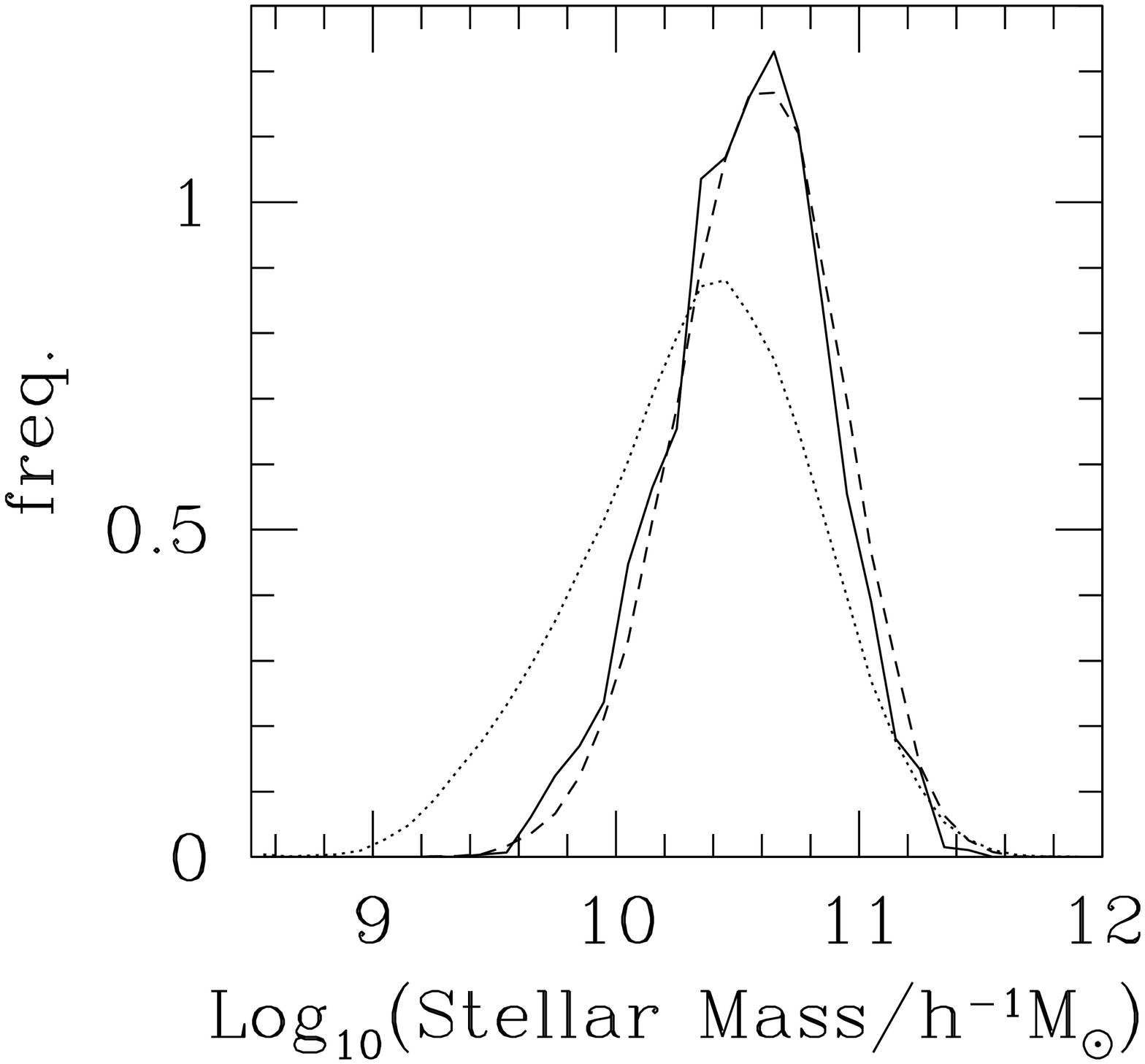,width=4.5cm}}
\put(295,-10){\psfig{file=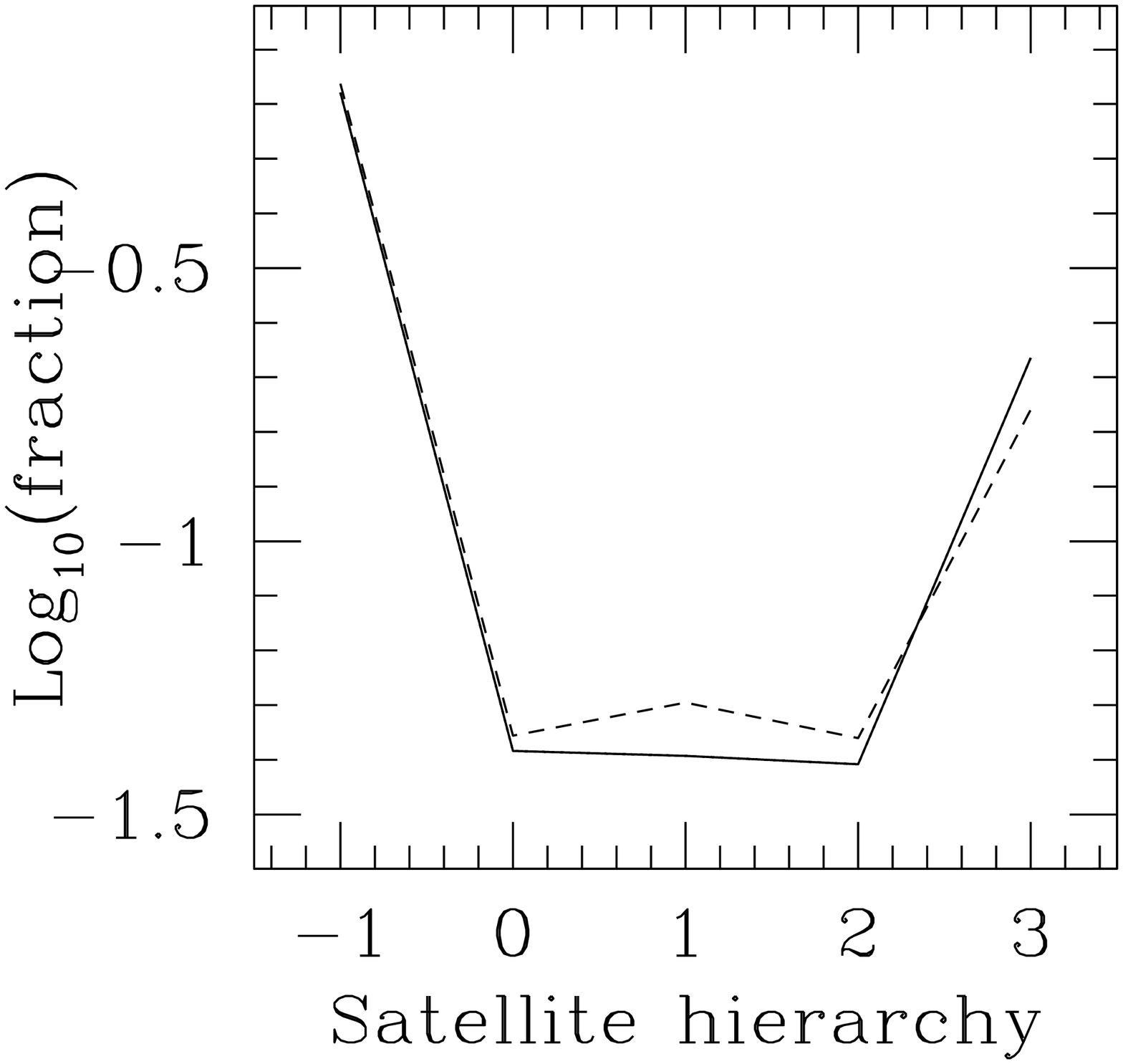,width=4.5cm}}
\end{picture}
\caption{
Comparison between AGN host (dashed lines), control galaxy (solid lines), and
full galaxy population properties (dotted lines), indicated in each panel (the distributions
are normalised to ensure unit integrals).  The parameters
used to constrain the members of the control sample are the $g-r$ colour, the $M_r$ rest-frame
absolute magnitude, the redshift, the stellar mass, the local density (in this subpanel
the grey long-dashed line corresponds to the high OIII luminosity AGNs),
the luminosity of the host group, and the fractions of central galaxies, brightest, second
brightest and faint satellites (corresponding to satellite hierarchies of $0, 1, 2$ and $3$).
A satellite hierarchy of $-1$ indicates galaxies not associated to any group { with masses
above the chosen minimum mass threshold}.
The distributions of distance to the nearest cluster of galaxies (in units of virial radii)
is also shown but is not used to define membership for the control sample.
}
\label{fig:control}
\end{figure*}

{  Due to the flux limit affecting the galaxy sample from which the groups are identified,
the number of galaxies per group does not correlate well with mass.}
We use the virial theorem to compute the virial mass of groups which is given by
\begin{equation}
 M_{vir}=\frac{3\sigma_{v}^2 R_{vir}}{G}
\end{equation}
where $\sigma_{v}$ is the line-of-sight velocity dispersion and $R_{vir}$ is estimated 
as in Merch\'an \& Zandivarez, (2005),
\begin{equation}
 R_{vir}=\frac{\pi}{2} \frac{N_{g}(N_{g}-1)}{\Sigma_{i>j}R_{ij}^{-1}}
\end{equation}
where $N_{g}$ is the number of galaxy members and $R_{ij}$ are the
relative projected distances between galaxies.
{  In order to ensure a high completeness for our sample of groups, 
we apply a second cut in group mass of $M>3\times10^{12}$h$^{-1}M_{\odot}$.}

We divide our sample of groups in (i) a Cluster Sample, comprising
all groups with virial masses $M_{vir}>10^{14}$h$^{-1}M_{\odot}$, to be used
as landmarks of high density regions for a global density estimator,\footnote{
The choice of minimum mass can be varied with only minor effects on our conclusions.
We tested limits as low as $M_{vir}>10^{13}$h$^{-1}M_{\odot}$ consistent with
the lower limit adopted by Gonz\'alez \& Padilla (2009).
} and (ii)
the rest of the groups, called Hosts Sample; a proxy for their masses will be used as indicators
of the immediate environment of galaxies.\footnote{In Gonz\'alez \& Padilla (2009) it
was shown that for fixed local density environments, most of the remaining global variation in 
galaxy properties were due to variations in the host halo mass; the dependencies
shown by the galaxy properties after fixing their local density and 
host halo mass could be due to the assembly history of
their host haloes (a minor effect).}

The Cluster Sample is used to define the global environment of galaxies as follows.
For each galaxy we compute the distance to all clusters in units of the cluster virial
radius.  The shortest distance is used to tag each galaxy so that samples at different
distances from clusters can be constructed using this parameter.  
In the case where a galaxy falls within
one virial radius of a cluster, in projection, and its velocity difference is $\Delta v<500$km/s,
we consider the galaxy to lie within the cluster.  The velocity difference is adopted to
take into account the ``finger-of-god" effect.  

We use the Hosts Sample to assign individual galaxies a host dark-matter halo mass.  
If a galaxy lies within
a group - if it has a host group or host dark-matter halo - 
its virial mass is assigned to the galaxy.  Galaxies
outside groups are assumed to be hosted by groups of masses below the {  completeness limit of
the catalogue (groups with four or more members)}, $M_{host}<3\times 10^{12}$h$^{-1}M_{\odot}$.  
In turn, once all the galaxies are assigned
to groups, a Group Luminosity is measured for each group in the sample by summing the luminosities of the $4$ 
brightest galaxy members in each group.  This quantity is a better proxy for the
true, underlying group mass than the virial estimate (Eke et al. 2004; Padilla et al. 2004), and it is
our parameter of choice to place further cuts on our samples of galaxies
{ (the stellar mass of member galaxies could be a better proxy
for the group mass than the galaxy luminosity; e.g. More et al., 2010).}
Out of our full samples of DR7 galaxies and DR4 AGN hosts, $66$ and $69\%$ reside in groups above the
completeness limit, respectively.

\subsection{Control samples for AGN hosts}

Given the possible selection biases affecting the AGN hosts, we
construct samples of control galaxies by randomly selecting those with no detected AGN activity
in the SDSS-DR4, such that they reproduce (to a $\sim 10\%$ or better) the normalised distributions
of redshifts, absolute magnitudes (rest-frame r-band), {  stellar masses,} local density $\sigma_5$, 
and $r_{90}/r_{50}$ r-band concentrations
of the AGN hosts.  Additionally, we require that the distributions of host group luminosity
are the same between control and AGN samples.  In the case of AGN hosts with no associated
groups, we force the control samples to have the same fraction of such objects.
{  The control samples also contain the same fractions of central, 
brightest, second brightest and faint satellites, than the AGN hosts; in our case
the brightest galaxy (r-band) in the group is considered the central galaxy.}
This procedure is similar to the one adopted by P\'erez et al. (2009) to study
the variations of galaxy properties as a function of environment when these are in pairs. 
As we are interested in the variations of galaxy properties when these host an AGN,
{  our selection of the control sample
uses the stellar masses as well;
for the general galaxy population this is not expected to produce
important changes after the luminosity, local density, and host halo mass
are constrained (P\'erez et al. 2009).}  P\'erez,
Tissera \& Blaizot (2009) claim that an adequate control
sample can help disentangle morphological and environmental effects on the galaxy population.

As we will show later in this work, the distance to the nearest cluster of galaxies
produces only mild changes in the galaxy properties, and therefore we have not
used this parameter to place further constraints on the control samples.

\begin{figure*}
\begin{picture}(430,395)
\put(0,-15){\psfig{file=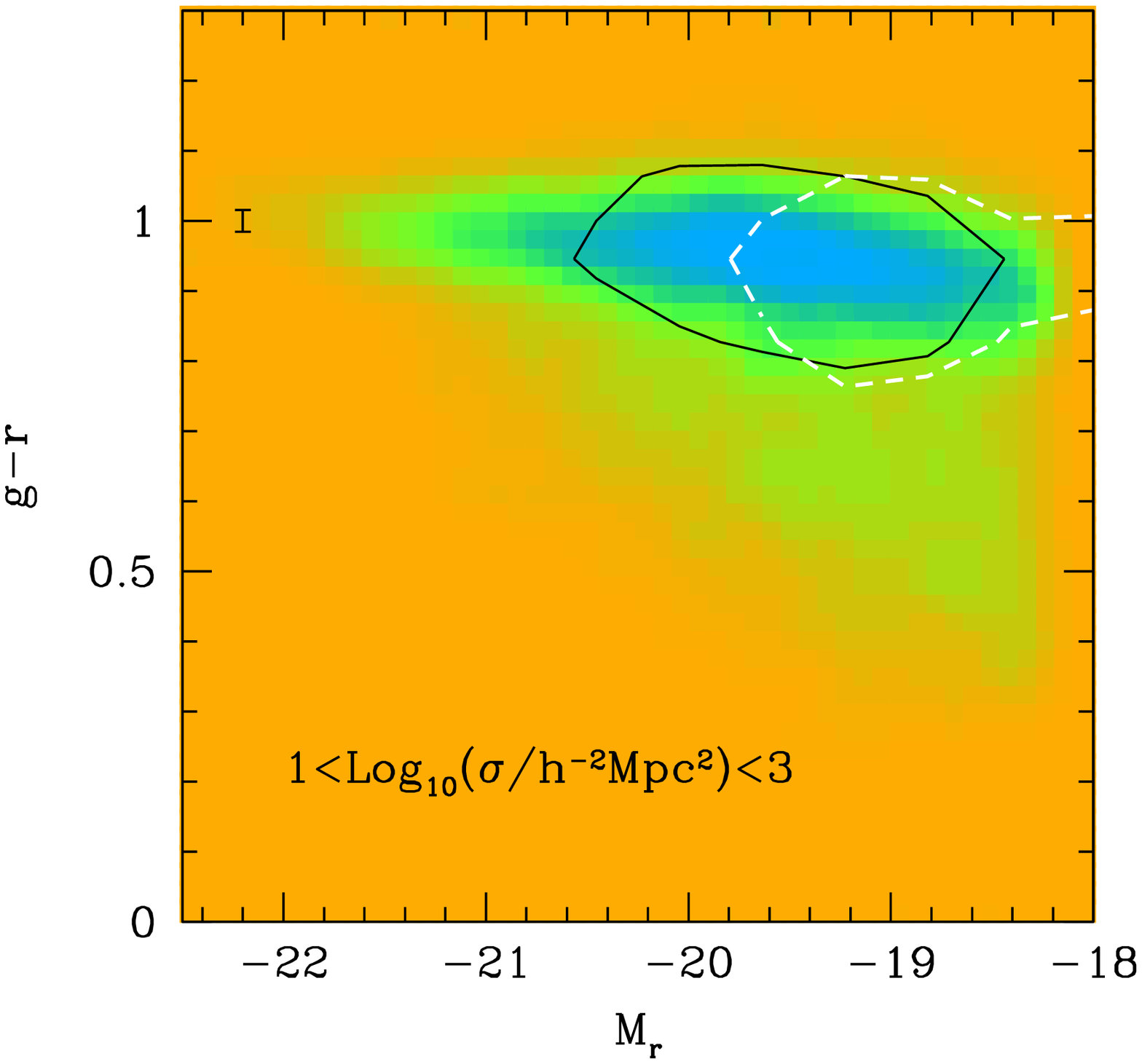,width=8.cm}}
\put(220,-15){\psfig{file=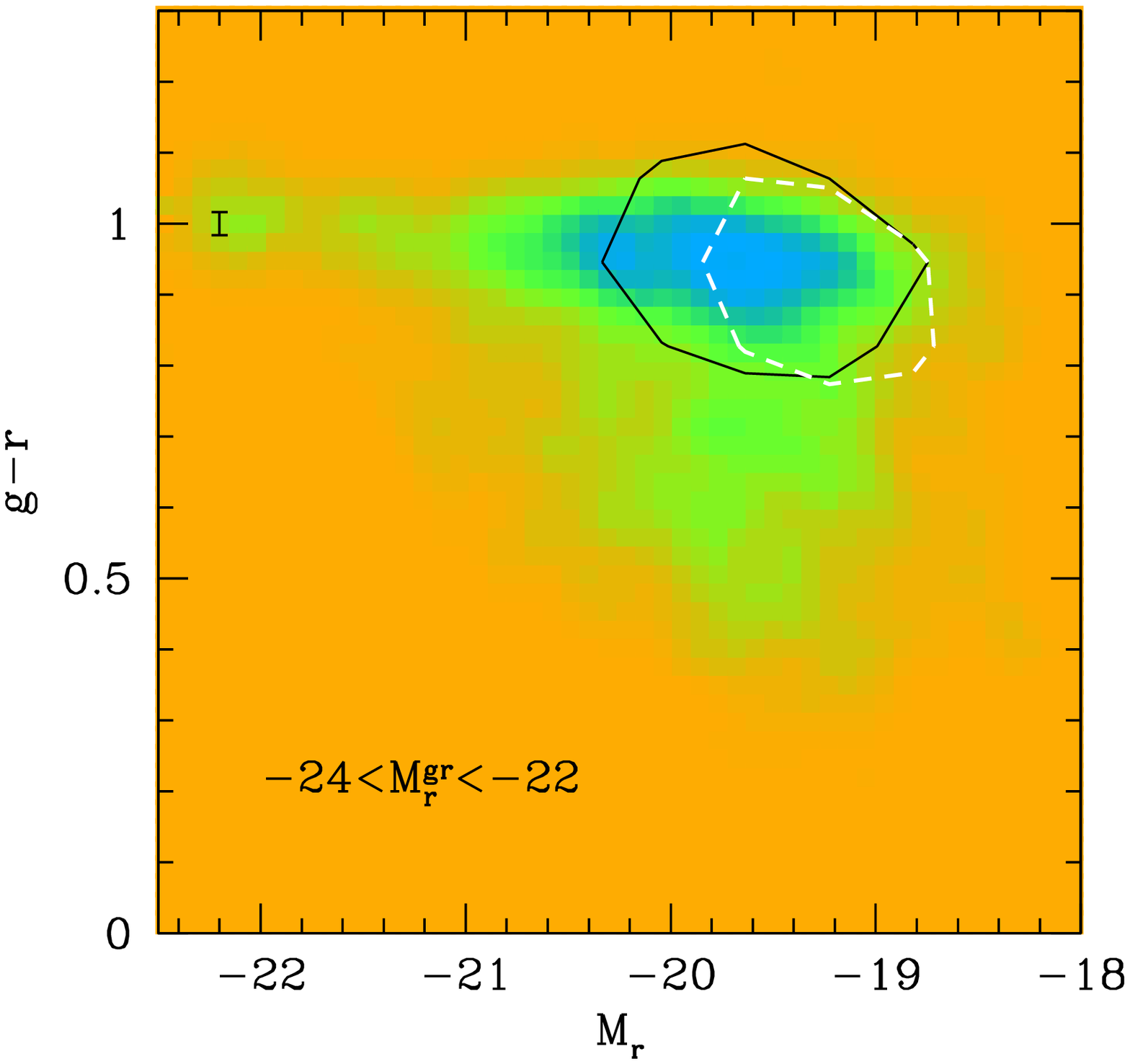,width=8.cm}}
\put(0,175){\psfig{file=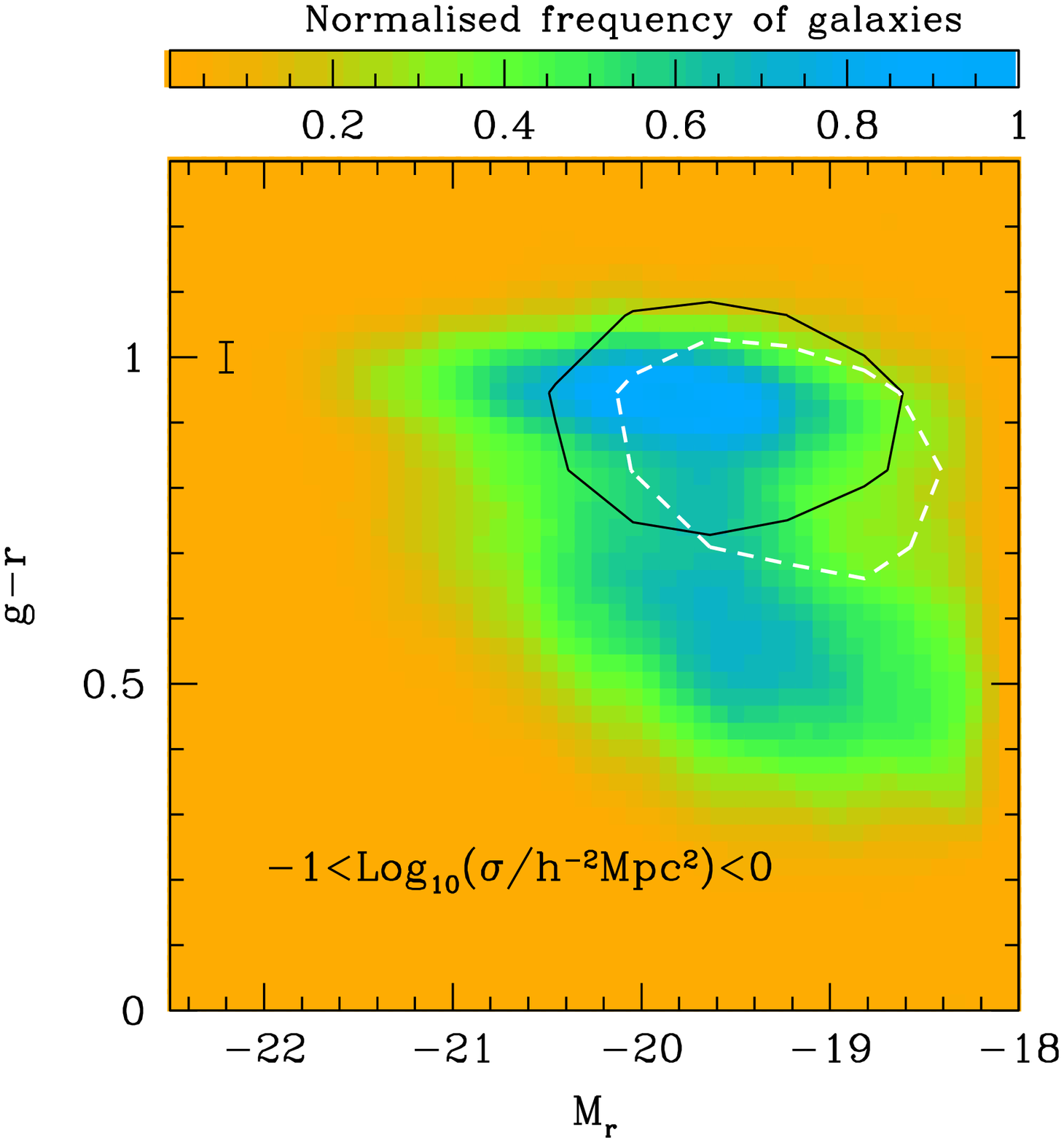,width=8.cm}}
\put(220,175){\psfig{file=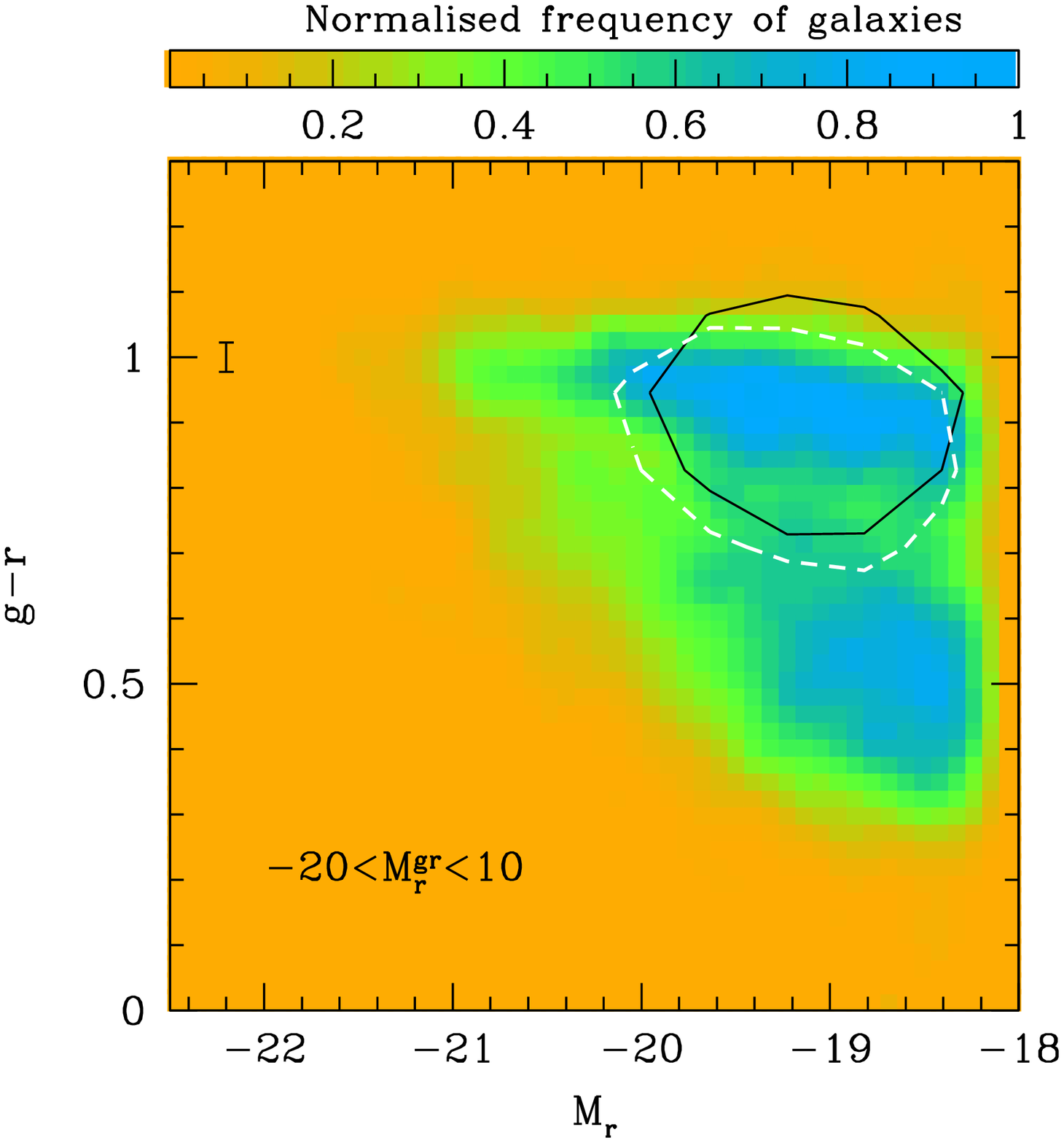,width=8.cm}}
\end{picture}
\caption{
Left panels: colour-magnitude relation for galaxies in low and high local density environments
(indicated in the figure, top and bottom panels, respectively).  
The frequency of galaxies (normalised to its maximum value) increases as the
colour progresses from yellow to blue.
Right panels: same as left panels but for different cuts on the group luminosity (used here
as a proxy for group mass).
The black solid and white dashed lines enclose $68\%$ of the AGN hosts and galaxies in the control sample for AGNs,
respectively.
}
\label{fig:cmr}
\end{figure*}

As a result of these constraints, of the full sample of non-AGN galaxies containing $130,828$
galaxies, a total of $7,311$ galaxies 
comprise the final control sample.  Figure \ref{fig:control} shows
in different panels the resulting normalised distributions of parameters used to define the control
samples.  Dashed lines correspond to the AGN hosts, solid lines to the control sample,
and dotted lines to the full galaxy population.  
As can be seen, the distributions
of properties for the AGN and control sample match each other in shape, even for the
distance to the closest cluster in units of virial radii which was not used to define the control
sample.  

By comparing the distributions for the full galaxy population (dotted lines) to that of
the AGN hosts, it can be seen that the latter are concentrated towards the red peak of
the colour distribution, and are characterised by higher luminosities (the peak of the
distribution of absolute magnitudes is about one magnitude brighter) and stellar masses
(consistent with e.g. Silverman et al. 2009)
and therefore higher redshifts.  It can be seen that the normalised distribution of host group
luminosities is lower in amplitude for the full sample of galaxies at this range of luminosities.  This
is due to a higher fraction of galaxies with no host group in comparison to the AGN host
samples, also consistent with Silverman et al. (2009) who find that AGN hosts are preferentially found
in groups of higher mass.  Also, the fraction of central galaxies in the full sample is $(14.0\pm0.1)\%$ compared
to a higher fraction of $(16.4\pm0.4)\%$ in the AGN sample (as in Pasquali et al. 2009). It 
should be borne in mind that these differences may be due, at least in part,
to selection effects associated to the detection of AGN features.  The local density and
the distance to the closest galaxy clusters show similar distributions for the full and AGN
host samples {  indicating constant fractions of AGN as a function of these two environment
measurements.  The grey long-dashed line in the lower-left panel shows the distribution of local
densities for high OIII luminosity AGN
{ ($Log_{10}(L_{OIII}/L_{\odot})>6.4$, corresponding to the median luminosity of the sample and the LINER
limit)}, which shows a lower abundance of bright AGNs at higher
densities, in agreement with the results by Kauffmann et al. (2004) and Popesso \& Biviano (2006)}. \footnote{
To provide the statistical significance of this result 
we calculated the fraction of AGN hosts located in local density environments
characterised by $\sigma_5>10$h$^{-2}$Mpc$^2$, and find $(8.0\pm0.2)\%$ for bright AGN, and $(9.6\pm0.2)\%$ for the
full AGN sample, indicating a $>5\sigma$ detection of this effect.}

  Finally, a word of warning should be put in with respect to the BH population of the galaxies that comprise
the control sample.  Some unknown fraction of these galaxies could harbour dormant BHs, and their distribution
of masses could well reproduce that of the AGN sample (even if the distributions of concentrations
are similar).  However, there could also be a population
of galaxies with central black holes of completely different masses (even zero mass).  We will bear
this in mind when comparing the properties of the AGN hosts and galaxies in the control sample.

\section{Galaxy sequences and the color-concentration diagram}

In this section we show results on one- and two-dimensional distributions of
galaxy properties (colours, bulge concentration and galaxy luminosities), for
samples of galaxies limited by intrinsic luminosity, concentration, 
host halo mass, local density, distance to the nearest cluster of galaxies, and by whether 
they present an AGN.
Our sample is not volume limited since it comprises galaxies in the SDSS-DR7 out to $z=0.1$, with
$-23<M_r<-18$.  We correct for this incompleteness by using a $1/V_{MAX}$ weighting
scheme (as in Padilla \& Strauss 2008), bearing in mind the possible effect of large-scale structure
on our results not taken into account in the weighting procedure.

\begin{figure*}
\begin{picture}(430,410)
\put(0,-40){\psfig{file=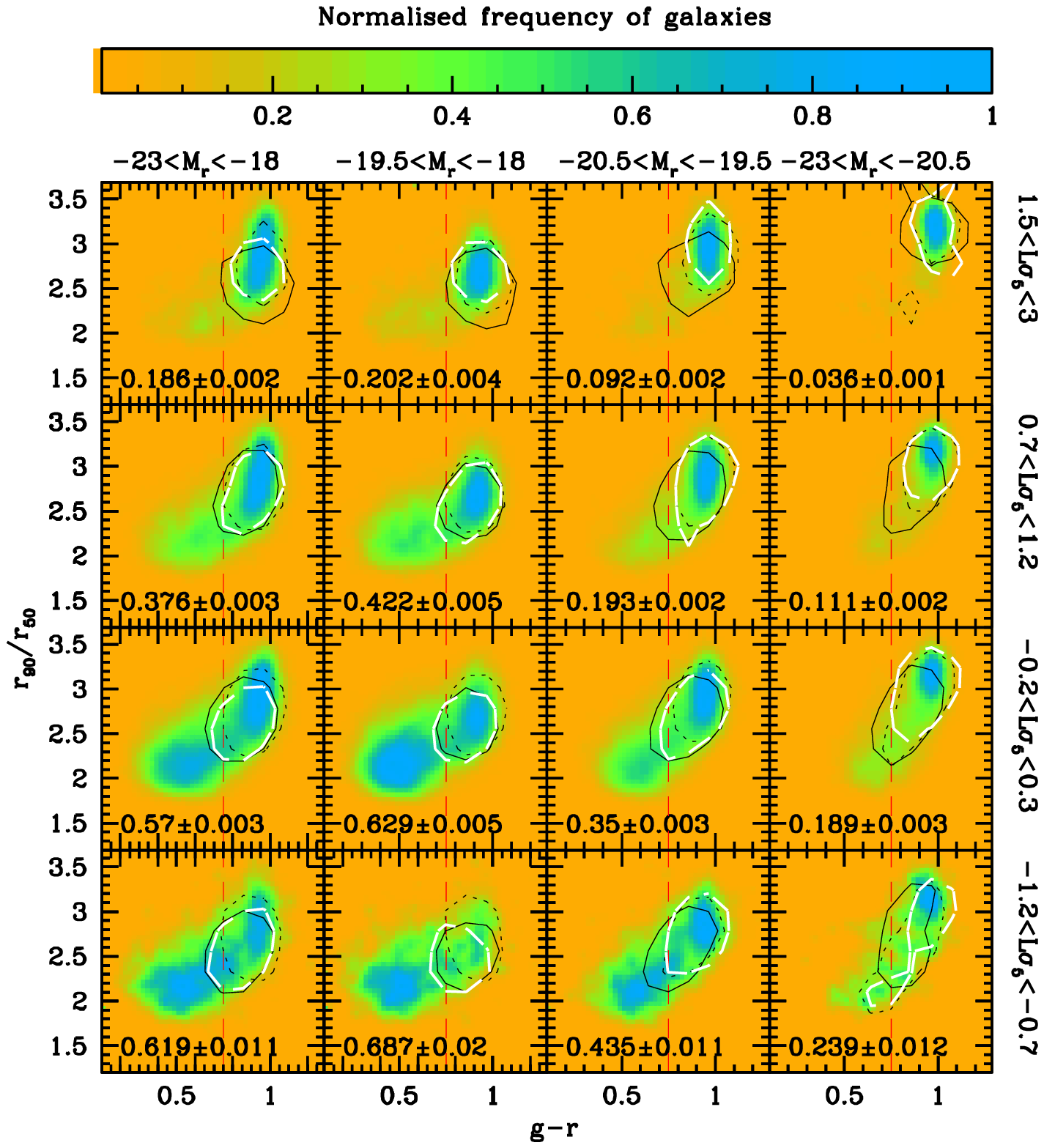,width=18.cm}}
\end{picture}
\caption{
Distribution ($1/V_{MAX}$ weighted) of galaxies in the $r_{90}/r_{50}$ vs. $g-r$ colour
plane, for different galaxy luminosities (ranges indicated on top of the upper sub-panels) and
located in regions with different local densities (ranges indicated on the right, $L\sigma_5$
is shorthand for $\log(\sigma_5/h^{-2}Mpc^2)$).  The
frequency of galaxies increases as the colour progresses from yellow to blue.  Lines enclose
a $68$ percent of the population of OIII bright (solid) and faint (dashed) AGN galaxies. 
The white dashed line shows the same contour for the full AGN control sample.
The dashed red lines indicate the value of $g-r$ used to compute the fraction of 
blue galaxies shown near the bottom of each panel.
}
\label{fig:paleta}
\end{figure*}

We will follow previous works by 
Gonz\'alez \& Padilla (2009) and Ceccarelli, Padilla \& Lambas
(2008), and define two proxies for galaxy environment.  (i) The local density, $\sigma_5$, 
designed to follow local effects such as galaxy-galaxy interactions, measured
using the cylindrical volume with radius equal to the distance to the fifth nearest neighbor.
The length of the cylinder corresponds to a redshift difference of $500$kms$^{-1}$,
and neighbors are required to satisfy $M_r<-20.5$ (neighbors are drawn from a 
volume limited subsample out to $z=0.1$); this method has been
widely used in the literature (Balogh et al. 2004; Baldry et al. 2005; among others).  
(ii) A quantity that is related to
larger scales and therefore to the history of mass build-up of the region where the galaxy
resides, rather than to the interaction history of the galaxy (modulo correlations between these
two quantities).
For this quantity, Gonz\'alez \& Padilla (2009) proposed to use 
the distance to the nearest cluster of
galaxies {  (in simulations)}, and Ceccarelli, Padilla \& Lambas (2008) used the distance to void centres
{  on SDSS-DR6 galaxies}. 
In both works, it is shown that the galaxy population shows variations as a
function of the global densities even when the local density is held fixed at
a given value.  For instance, Gonz\'alez \& Padilla (2009) showed that
at intermediate galaxy densities of $-0.5<\log(\rho/10^{10}M_{\odot}/h^{-2}Mpc^3)<0.5$,
the fraction of red galaxies decreases from $0.45\pm0.10$ to
$0.21\pm0.02$ as the distance to clusters of galaxies increases from $d/r_{vir}<2$
to $d/r_{vir}>9$, where $r_{vir}$ correspond to the virial radii of the closest
cluster to any given galaxy.
This indicates that the local
galaxy density as traced by $\sigma_5$ does not necessarily probe the widest dynamical
range of the dependence of galaxy properties on environment due to residual global
variations {  (Also suggested by Weinmann et al. 2006, using the SDSS-DR2).  One
of the main aims of this paper will be to provide a firm statistical confirmation (or not)
of this effect, which was only marginally found by Ceccarelli et al.  Note that the
mass of the host halo { does not probe the widest dynamical
range in local environmental effects}, as it has been shown that the properties of galaxies in equal mass haloes 
depend on their assembly or SF history (Zapata et al. 2009).}  

\subsection{Colour-Magnitude relations}

We start our analysis by studying the colour-magnitude (CM) diagram for different
local density environments and different host group luminosities.
The CM diagrams for restricted samples of galaxies are shown
in Figure \ref{fig:cmr}.  The colour scale corresponds to the full galaxy sample,
and the black solid and white dashed lines to the AGN and AGN control samples, respectively.
The left panels correspond to 
two different ranges of local density and the right panels to galaxies in groups of different
total luminosity measured in the r-band ($M_{gr}^r$, used as a proxy for group mass).  
In both cases there is a 
clear red-sequence which becomes tighter for higher galaxy luminosities.
This sequence has been thoroughly used as a tool to detect clusters at both
low and high redshifts (e.g. Gladders \& Yee 2005).  We note that
we obtain similar CM diagrams for low densities and low host group luminosities (shown on the right), and
for high densities and high group luminosities.  In the former there are clear blue clouds,
which almost completely disappear for the latter.  This behaviour is well documented and
can be explained via
the density-morphology relation (e.g. Dressler 1980; Postman \& Geller 1984; Whitmore
\& Gilmore 1991; Goto et al. 2003; Postman et al. 2005).  The individual errorbars on the left 
of each panel show the $68\%$ width of the red sequence; it is clear that for higher
densities and higher host group luminosities, the red sequence is more sharply defined (has a 
smaller dispersion).

Both, AGN hosts and galaxies in the AGN control sample, show wider distributions
in colour than the red-sequence shown by the full sample of galaxies due to the effect noticed
earlier when inspecting Figure \ref{fig:control}, that the AGN host colour distribution
is similar to the red part of the bimodal colour distribution of the full galaxy
sample, although slightly wider and shifted towards bluer colours.  
We note that the width of the distributions of AGN colours does not
change significantly when the density or the host group luminosity increase (there is
a marginal narrowing of the AGN colour distribution at higher densities).  
{  AGN hosts show a slight tendency to become redder and brighter
at higher local densities (a correlation in agreement with a colour-magnitude
relation also present in AGN hosts).} 
Since
this effect is also visible for the control samples, it must be a result of the
morphology and environment of the AGN hosts.

When comparing AGN hosts and control samples, there are some noticeable differences 
between them.  
Regardless of the local density and host group luminosity, the hosts of AGN tend
to be slightly redder, compared to galaxies in the control sample; this will be confirmed in the
following subsections.
As a function of local density, it can be seen that the AGN hosts become brighter 
relative
to control galaxies as the density increases.  This is more clearly seen as a function
of host group luminosity.  These two results are 
consistent with a picture in which 
AGN are hosted by larger galaxies in clusters, which { are likely to host higher mass BHs and also}
are more efficient in retaining
gas in spite of the adverse intracluster conditions in comparison with low mass objects
which are likely to be gas stripped;  we remind the reader that the proportion of
central and satellite galaxies are the same for the samples of AGN hosts and their control galaxies.

\subsection{Dependence on local density}

With the aim of understanding the changes in the galaxy population in different
large-scale environments, we measure the variations in the galaxy
sequences in a $g-r$ colour vs. bulge concentration diagram.
This is shown in Figure \ref{fig:paleta}, where in the frequency scale 
bluer colours indicate a large concentration
of objects ($1/V_{MAX}$ weighted) and yellow shows regions with a lack of galaxies.  
The left panels show galaxies with no restriction on luminosities 
($-23<M_r<-18$), and low, intermediate and
bright galaxies are shown towards the right; higher local densities, $\sigma_5$, are shown
towards the top panels.

The plots show two distinct populations, one containing blue, low bulge concentration galaxies,
and another with red galaxies with a wider range of bulge concentration values.
The red galaxy population shows an elongated sequence in this figure,
which due to the colour-magnitude relation appears slightly narrower in colour when the range of intrinsic
luminosities is restricted.  
The red long-dashed lines in the figure correspond to 
$g-r=0.75$ which is 
used to divide the population into blue and red galaxies.  The number at the bottom
of each panel shows the fraction of blue galaxies (the error corresponds to Poisson
uncertainties).
For the left panels, which consider the full intrinsic luminosity range, there is a clear
trend of a lower fraction of galaxies in the blue, low bulge concentration population
as the local density increases.  As can be seen on the columns to the right, this
effect is present at all galaxy luminosities.
These results are in agreement with a number of studies of the dependence of galaxy properties
on the environment including Balogh et al. (2004), Baldry et al. (2006), and {  Weinmann et al. (2006)}.  Consistent 
results were also obtained for galaxies in voids (low local densities), which were found to be bluer
and with higher star-formation activity than galaxies in average environments
(e.g. Goldberg et al. 2005; 
Padilla, Ceccarelli \& Lambas 2005; Rojas et al. 2005; Hoyle, Vogeley \& Rojas 2005;
Ceccarelli et al. 2006; Ceccarelli, Padilla \& Lambas 2008); and for clusters
of galaxies on the high local density end, where galaxies show the well-known
cluster-centric radius-morphology relation (e.g. Dressler 1980; Postman et al. 1995;
Dom\'\i nguez et al. 2002). 

\begin{figure*}
\begin{picture}(340,360)
\put(-80,-45){\psfig{file=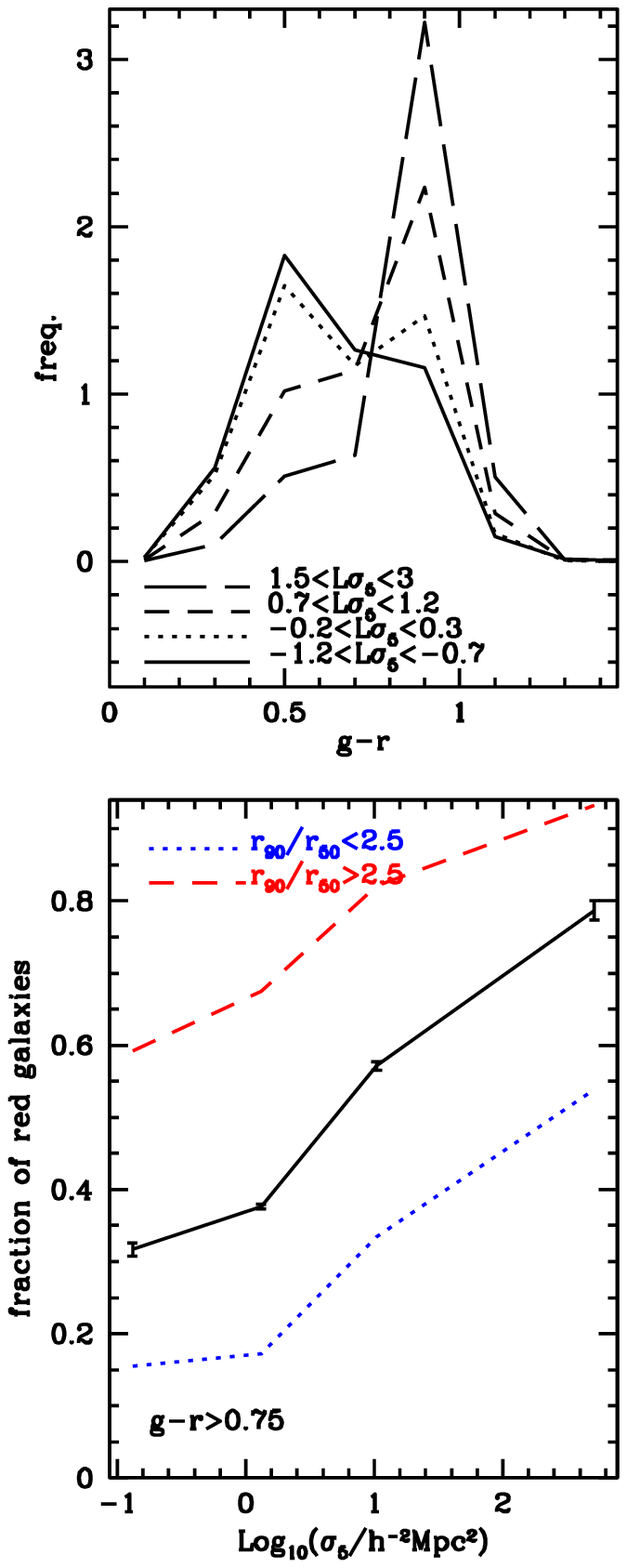,width=15.cm}}
\put(-40,-45){\psfig{file=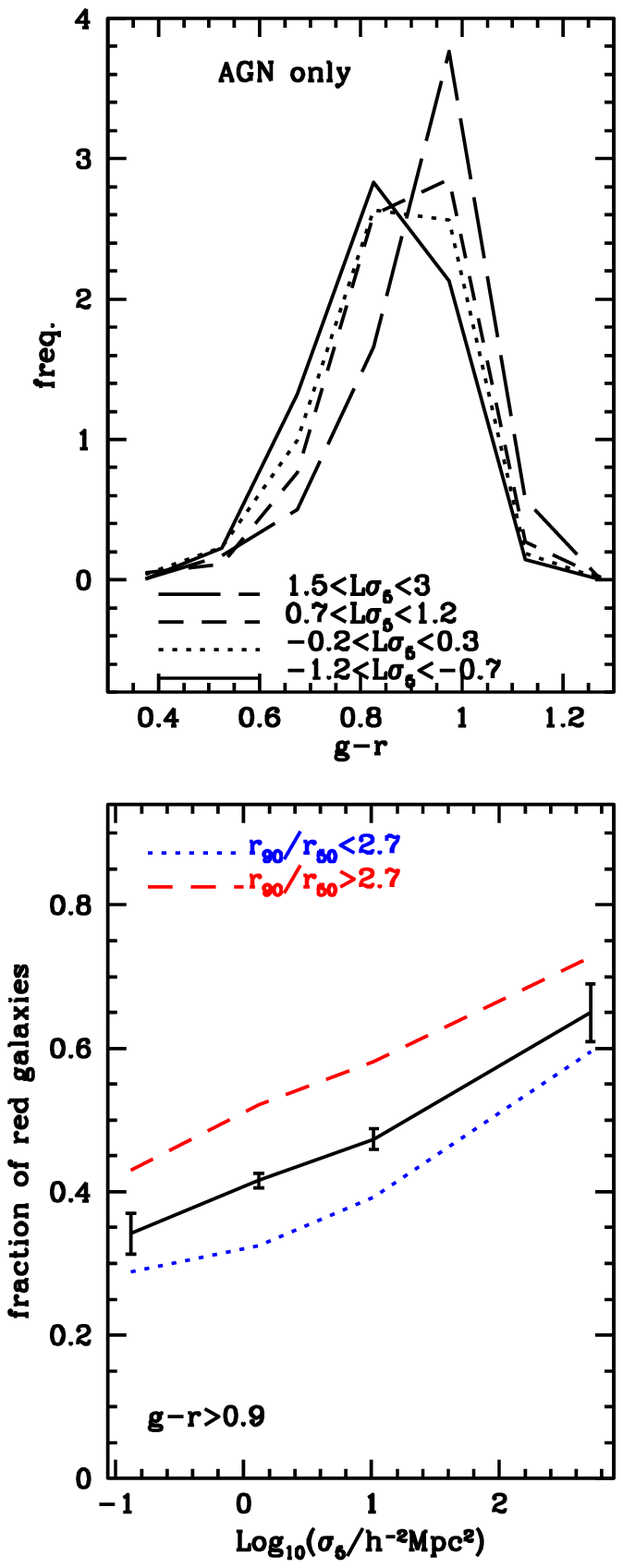,width=15.cm}}
\put(120,-45){\psfig{file=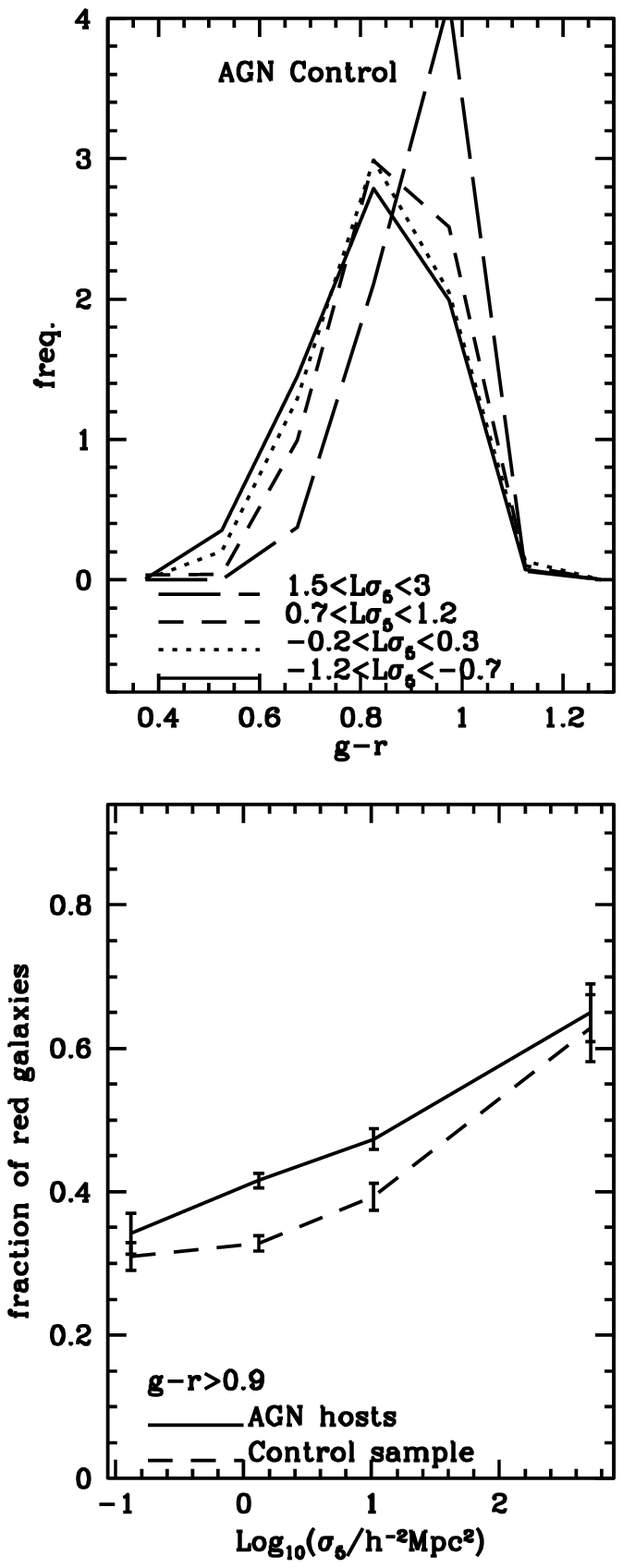,width=15.cm}}
\end{picture}
\caption{
Top panels: $1/V_{MAX}$ weighted distributions of $g-r$ colour
for galaxies at different local densities $\sigma_5$, shown in different
line-types indicated in the figure key.  The left panel corresponds to non-AGN galaxies, the 
centre panels to the full AGN population, and the right panels to the AGN control sample.
Lower-left and lower-centre panels: fraction of red galaxies as a function of local density (solid lines)
regardless of concentration and for high (dashed) and low (dotted) concentrations.  The
cuts in colour and concentration are shown in the key.
The lower-right panel shows the fraction of red galaxies as a function of
local density for the control sample (dashed)
and for the AGN hosts (solid, repeated from the lower-centre panel).  Errorbars
show Poisson uncertainties on the mean.
}
\label{fig:rot}
\end{figure*}

{ Our study of the environmental dependence of AGN activity
will centre on the colours and concentrations
of AGN hosts instead of analysing the fraction of galaxies hosting AGN as done in
several works (Kauffmann et al. 2004; Martini, Sivakoff \& Mulchaey 2009).  Our results
will be comparable to those presented by Choi et al (2009, see also Waskett et al. 2005;
Gilmour et al. 2007; Coldwell et al. 2009; {  Pasquali et al. 2009}) who studied the morphological
types of the hosts.  }
The black line contours in Figure \ref{fig:paleta}
enclose $68$ percent of the AGN population in this diagram.  The solid
lines correspond to the brightest half (in {  extinction corrected} OIII luminosity) of the AGN sample; the dashed line 
to the galaxies with the fainter OIII emission half of the sample.  {  The median OIII luminosity of the AGN sample
is $Log_{10}(L_{OIII}/L_{\odot})=6.4$ (which roughly corresponds to the LINER limit).}
As can be seen, AGN in our sample tend to prefer hosts with colours redwards of the
blue population, in qualitative agreement with Choi et al. (2009) and also Schawinski et al. (2007).
It can also be seen that there are no strong trends in either the OIII bright or
faint AGN samples with the local density (for this reason we do not produce control samples
for AGN divided by OIII luminosity), {  except for a mild
tendency to reach lower host bulge concentrations and bluer host colours }in environments characterised
by lower local densities, {  for both high and low OIII luminosity AGN hosts.}
The white dashed lines show the regions where the AGN control samples lie in this diagram.  As
can be seen, most of the behaviour shown by the AGN comes from the morphology and
environment of their hosts; in comparison to the AGN hosts, particularly
those with low r-band luminosities, the control samples show
{  a more noticeable shift from low to high concentrations and $g-r$
colours as the local density increases; for instance for $-19.5<M_r<-18$ the
control sample is a good match to the high OIII luminosity AGN hosts at low densities, 
but a better match for the low OIII luminosity AGNs at the highest densities probed.}
We will come back to this point later in this section.  

The OIII bright AGN sample consistently
shows bluer colours (occupying the ``green" valley) {  and lower concentrations}
than the faint AGN.  This result can { 
be understood in terms of the effect described by Kewley et al. (2006),
who also find that LINERs (faint OIII AGNs) are more concentrated than
Seyferts (bright OIII AGNs), which they interpret as the latter being
likely associated with gas-rich disc galaxies where gas can supply SF and AGN activity
independently, whereas LINERs may have switched off their SF long ago and have maintained their
AGN status using gas supplied by mass loss from their stellar population or via mergers.  Another 
possibility comes from the work by Schawinski et al. (2007).}  They claim
that AGN are triggered by the same processes as the star formation (SF) in a galaxy, and
that the effect of AGN activity is that of shutting off the SF in
a galaxy.  Therefore, if the AGN activity is strong, i.e. the central accretion disc is
acquiring mass at high rates, the SF shutdown may still be in process, whereas once the
fuel has been consumed (low nuclear luminosity), the SF burst has already been quenched
by the previous high activity phase of the AGN and the galaxy appears redder.  This
is also in agreement with estimates of stellar ages (the time since the last SF episode)
by Schawinski et al., of more recent SF events for bluer AGN hosts. {  However,
it has been argued (e.g. Kewley et al. 2006) that other processes could be more likely the
source of the SF quenching, such as hot and cold gas stripping in satellites.}

\begin{figure}
\begin{picture}(230,420)
\put(10,-20){\psfig{file=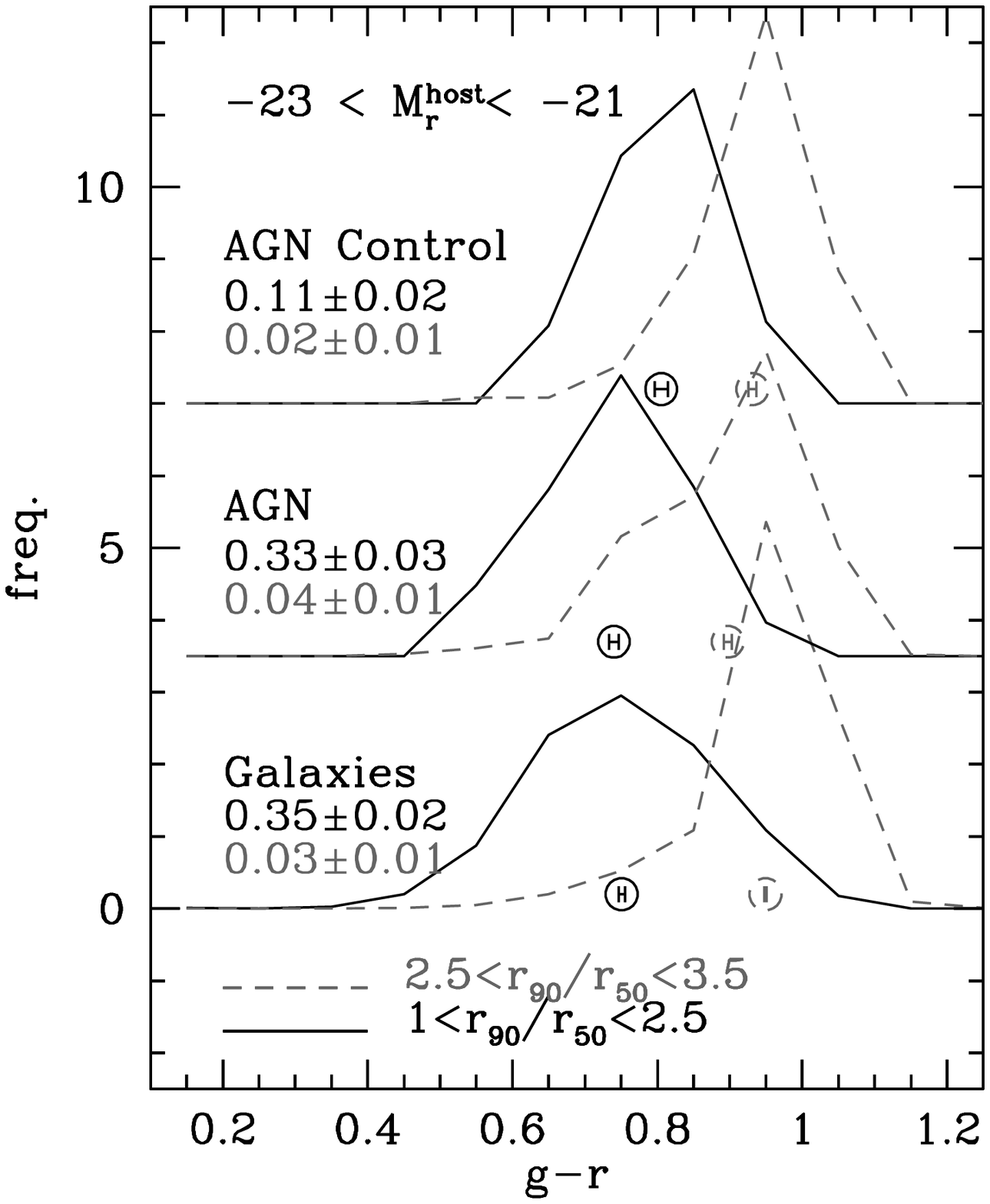,width=8.cm}}
\put(10,190){\psfig{file=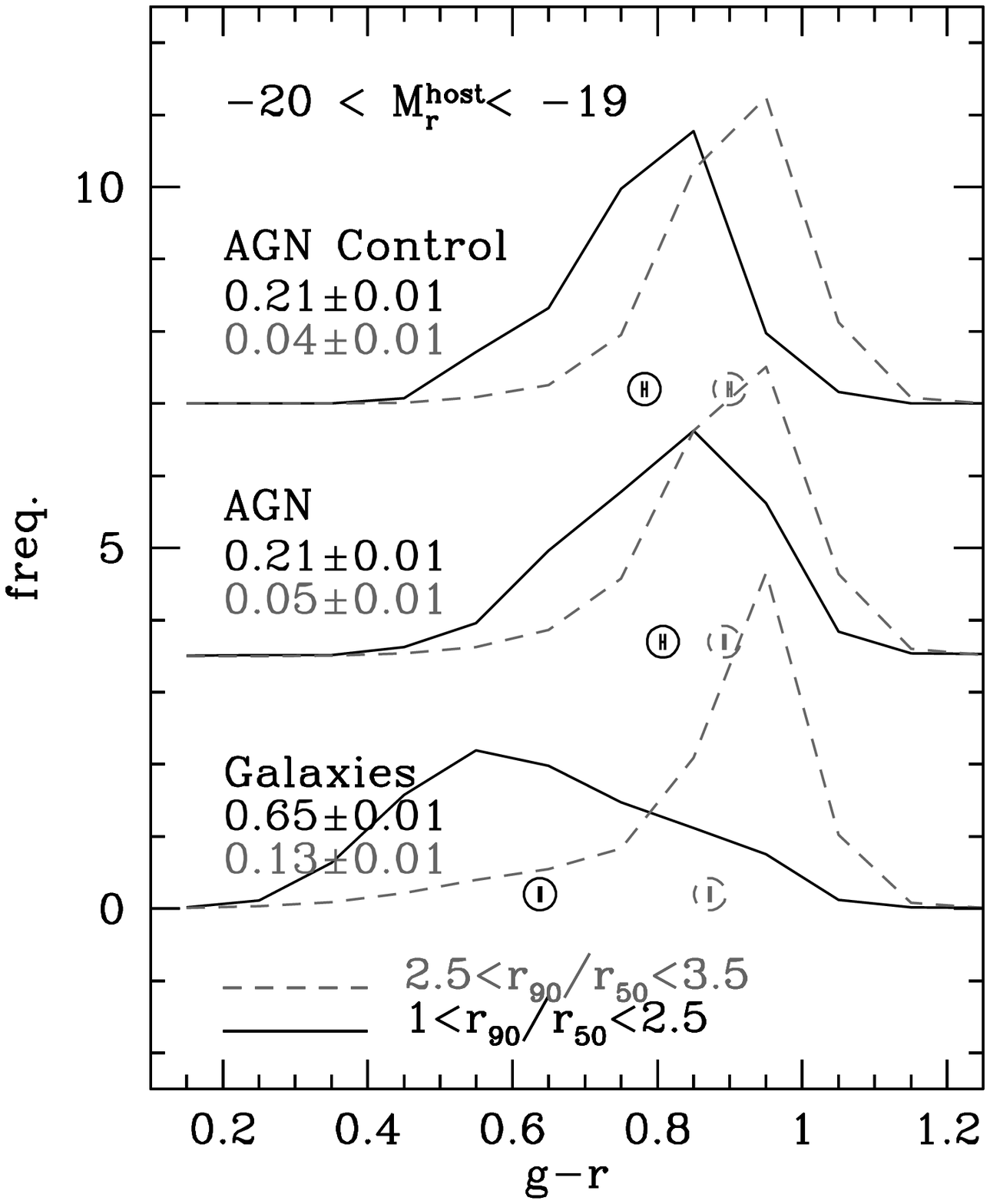,width=8.cm}}
\end{picture}
\caption{
Weighted distributions of $g-r$ colour for galaxies (lower lines),
AGN hosts (middle lines in each panel, displaced vertically by a constant value),
and the AGN control sample (upper lines, displaced further up vertically),
for low and high values of concentration (solid and dashed lines, respectively).
Top and bottom panels correspond to low and high galaxy luminosities.  The
luminosity and concentration ranges are indicated in the legend.
The circles show the average colour for the different samples (errorbars are the
Poisson error of the mean) and are used to allow a statistical comparison between the distributions (See
values in Table 1).
The fraction of blue galaxies is shown in black (grey) for the low (high) concentration sample.
This fraction is obtained using a limit of $g-r=0.75$ in all cases and are also shown in Table 1.
}
\label{fig:colagn}
\end{figure}

{  At the high galaxy luminosity (r-band) and low density end, there is some
evidence for a reversal of this relation, namely, high OIII luminosity AGNs tend to have higher $g-r$
colours and concentrations than low OIII AGNs, on average.  This is mainly due to the extension of the
sequence shown by the low OIII luminosity AGNs which tends to cover the full range shown by the general
galaxy population.}
This shows that the relation between AGN activity and star-formation
(as traced by galaxy colours) is not {  exactly} one-to-one; at the very least it depends
on host luminosity and concentration since
for fixed galaxy luminosity
and colour, we see that when the local density decreases the AGN activity 
shifts towards more extended, possibly disk-like objects.  The control sample shows the latter
behaviour independently of the local density.  We quantify these observations in Section 3.3.

The top-left panel of  Figure \ref{fig:rot} 
shows the $g-r$ colour distribution for galaxies with $-23<M_r<-18$ at
different local density environments (different line-types, indicated in the figure key).
The distributions show a clear bimodality with peaks located at roughly fixed positions,
with the red and blue peaks separated by $\Delta(g-r)=0.43\pm0.02$.
There is a clear shift towards the red galaxy population as the local density increases (consistent
with previous studies, e.g. O'Mill et al., 2009).
This trend is quantified in the lower left panel, where black dashed lines show
the fraction of red ($g-r>0.75$) galaxies as a function of $\sigma_5$ (errorbars correspond to
Poisson fluctuations
and are only shown for the sample with no restriction
on concentration to improve clarity).  We also show
this fraction in dotted lines for
low concentration galaxies, and in dashed lines for high concentration galaxies; as can be seen
the bluer colours of low concentration galaxies are also confirmed by these ratios.  

The top-centre panels of the figure show the colour distributions of AGN hosts, 
for the full range of
nuclear OIII luminosity.  As can be seen in the colour distributions, as well as in the red
fractions shown in the lower-centre panel, AGN hosts show a milder but
significant trend towards redder colours when embedded in higher density environments, as well
as higher red fractions as the concentration of the hosts increases.

\begin{table*}
\begin{minipage}{175mm}
\caption{
Average $g-r$ colour, fraction of blue galaxies ($g-r<0.75$), stellar masses ($LM^*$ is shorthand for 
$Log_{10}(M^*/$h$^{-1}M_{\odot})$) and fraction of central galaxies { (satellite hierarchies $-1$ 
and $0$)}, for the full galaxy sample, the AGN hosts,
and the AGN control sample, for two different ranges of r-band luminosities.  {  
The quoted errors correspond to the errors
on the mean, calculated using Poisson statistics.}
}
  \begin{center}\begin{tabular}{@{}cccccccccc@{}}
  \hline
{\tiny Sample }&{\tiny Luminosity }&{\tiny fr. blue gals. }&{\tiny fr. blue gals. }&{\tiny $\left< g-r \right>$ }&{\tiny $\left< g-r \right>$}&{\tiny $\left< LM^* \right>$ }&{\tiny $\left< LM^* \right>$ }&{\tiny fr. cent. }&{\tiny fr. cent. }\\
      & &{\tiny low $r_{90}/r_{50}$}&{\tiny high $r_{90}/r_{50}$}&{\tiny low $r_{90}/r_{50}$}&{\tiny high $r_{90}/r_{50}$ }&{\tiny low $r_{90}/r_{50}$}&{\tiny high $r_{90}/r_{50}$ }&{\tiny low $r_{90}/r_{50}$}&{\tiny high $r_{90}/r_{50}$ }\\
 \hline
{\tiny Control}&{\tiny $-20<M_r<-19$ }&{\tiny $0.21\pm0.01$ }&{\tiny$0.04\pm0.01$}&{\tiny$0.782\pm0.003$ }&{\tiny$0.900\pm0.002$}&{\tiny$10.37\pm0.30$}&{\tiny$10.43\pm0.23$}&{\tiny$0.74\pm0.03$}&{\tiny$0.70\pm0.02$}\\
{\tiny AGN }&{\tiny $-20<M_r<-19$}&{\tiny $0.21\pm0.01$}&{\tiny$0.05\pm0.01$}&{\tiny$0.808\pm0.003$ }&{\tiny$0.893\pm0.002$}&{\tiny$10.38\pm0.26$}&{\tiny$10.47\pm0.17$}&{\tiny$0.58\pm0.02$}&{\tiny$0.59\pm0.01$}\\
{\tiny Full}&{\tiny $-20<M_r<-19$ }&{\tiny $0.65\pm0.01$}&{\tiny$0.13\pm0.01$}&{\tiny$0.638\pm0.001$ }&{\tiny$0.873\pm0.001$}&{\tiny$10.16\pm0.08$}&{\tiny$10.39\pm0.09$}&{\tiny$0.623\pm0.006$}&{\tiny$0.513\pm0.005$}\\
  \hline
{\tiny Control}&{\tiny $-23<M_r<-21$ }&{\tiny $0.11\pm0.02$}&{\tiny$0.02\pm0.01$}&{\tiny$0.805\pm0.009$}&{\tiny$0.931\pm0.005$}&{\tiny$11.07\pm1.23$}&{\tiny$11.09\pm0.64$}&{\tiny$0.90\pm0.10$}&{\tiny$0.86\pm0.05$}\\
{\tiny AGN }&{\tiny $-23<M_r<-21$}&{\tiny $0.33\pm0.03$}&{\tiny$0.04\pm0.01$}&{\tiny$0.740\pm0.007$}&{\tiny$0.897\pm0.005$}&{\tiny$11.06\pm0.72$}&{\tiny$11.17\pm0.55$}&{\tiny$0.87\pm0.06$}&{\tiny$0.80\pm0.04$}\\
{\tiny Full}&{\tiny $-23<M_r<-21$ }&{\tiny $0.35\pm0.02$ }&{\tiny$0.03\pm0.01$ }&{\tiny$0.751\pm0.004$}&{\tiny$0.949\pm0.002$}&{\tiny$11.05\pm0.45$}&{\tiny$11.18\pm0.24$}&{\tiny$0.84\pm0.03$}&{\tiny$0.81\pm0.02$}\\
\hline
\label{table:1}
\end{tabular}
\end{center}
\end{minipage}
\end{table*}

The right panels show these results for the AGN control sample.  As can be seen, the
colour distributions and their variation on local density is qualitatively similar to
that shown by the AGN hosts.
The lower right panel shows in a solid line the fraction of red galaxies for the
AGN sample (repeated from the lower-centre panel) and in dashed lines that of the control sample.
The variation of this fraction over the three orders of magnitude in local density shown in the figure, is equal in amplitude
for the control galaxies and AGN hosts ($0.32\pm0.02$ and $0.31\pm0.01$ increases in red fractions,
respectively; Poisson errors) and lower than for the full galaxy sample 
(a red fraction variation of
$0.58\pm0.01$).  A significant difference between
galaxies in the control sample and AGN hosts is that the former
are characterised by {  a lower average red fraction 
of $0.415\pm0.014$ and $0.470\pm0.013$, respectively (with a $2.9 \sigma$ significance 
level).}
Bearing in mind the black-hole vs. bulge mass relations (e.g. Magorrian et al. 1998; Haring \& Rix 2004),
this difference in colours may be due to the presence of larger bulges in the AGN hosts (which
on average will have higher mass black-holes at their centres in order to have been detected as AGN).
{  Alternatively, as the control sample may include galaxies with BHs with similar masses
as the AGN hosts, but in their dormant phase, this could also point to a recovery of the SF activity
in the control sample after their last AGN cycle; this last speculative interpretation would also
allow to understand that this difference is not significantly detected at the highest local 
densities due to the higher difficulty for gas cooling in such environments.}

Several works on the dependence of the AGN population on environment study
the fraction of galaxies that host active nuclei.  For example, in
a combined study of high and low redshift galaxies using DEEP2 (Faber et al. 2007) 
and SDSS, respectively, Montero-Dorta et al. (2009) find that galaxies in the SDSS 
show a decreasing fraction of AGN in the red sequence towards high density environments. Our
result is complementary to this approach, indicating that regardless of the fraction of
active galaxies, their hosts show a slight tendency to become redder as the local density 
increases.

Later in this section we will study
whether the colours of galaxies show any dependence
when the local density is held fixed and the distance to clusters is
allowed to vary.

\subsection{AGN and galaxy sequences}

In this subsection we quantify the behaviour of AGN and how it compares to that of
their control sample and the full galaxy population at different local densities, and for
different intrinsic host galaxy luminosities.   Pasquali et al. (2009) indicate
that the AGN activity is more likely to be found in central galaxies; in the following
comparisons we will verify whether the fraction of central galaxies of subsamples of the full, control
and AGN populations under comparison are similar, as we will also do with their average stellar masses.
{ In order to estimate the total number of central galaxies in our samples we use the satellite
hierarchy shown in Figure \ref{fig:control}, which is equal to $0$ if a galaxy is central in a
SDSS group in the hosts sample (which comprises groups with masses 
$M>3\times10^{12}$h$^{-1}M_{\odot}$), hierarchy $>0$ if a galaxy is a satellite in one of these groups, and $-1$ if
no group is associated to the galaxy.  Therefore, the sample of galaxies with satellite hierarchy $0$
does not represent the full central galaxy population.  Zehavi et al. (2004) use the Halo Model
(Cooray \& Sheth, 2002)
to measure the minimum
dark-matter halo mass from which haloes start to contain satellites brighter than $M_r=-21$, and find it to 
be 
$M\simeq10^{13}$h$^{-1}M_{\odot}$.  On the other hand, Zheng et al. (2005) extend the study
of the Halo Occupation model to lower luminosities using models of galaxy formation.  Combining
their results with those by Zehavi et al., we expect that
less than $5\%$ of $-21<M_r<-19$ galaxies
in haloes below the lower mass limit of the hosts sample
will be satellites.  Therefore, from this point on,
 we will consider as central galaxies those with satellite hierarchies of $0$ and $-1$ (for $M_r<-19$).}

Figure \ref{fig:colagn},
shows the colour distribution of high and low concentration galaxies and AGN hosts,
in two different luminosity ranges. 
{  As can be seen, galaxies selected this way show almost exclusively unimodal
colour distributions which will allow us to study average colours for each subsample.}
The results for objects with $-19.0 < Mr< -18.0$  
and $-23.0 < Mr < -20.5$ (top and bottom panels, respectively) clearly show 
that faint AGN hosts show more similar fractions of blue galaxies between high and low
concentrations than high luminosity AGN hosts.  In the case
of no restriction on AGN activity, galaxies with similar luminosities show a similarly strong
behavior of much lower values of $g-r$ for low concentrations regardless of galaxy luminosity.
These results are supported by the fractions of blue galaxies and blue AGN hosts
(shown in the panels, and in Table 1),
and the average $g-r$ colours for each of the populations studied in the figure (shown
as circles with errorbars, dashed and solid for high and low concentrations, and in Table 1).
Namely, we find a clear difference between the SF activity (traced
by the $g-r$ colour) and that of AGN, indicating that
low concentration (disc-like) hosts are characterised by bluer colours for AGN hosts of
higher luminosities.
Instead, the control AGN samples (and normal galaxies) show redder colours for higher galaxy luminosities.
{ With respect to AGN hosts, control galaxies show
redder colours (lower fractions
of blue galaxies).  
It is possible that control galaxies contain similar BH masses in a dormant phase such
that their SF activity has been quenched during the previous cycle of activity.  One
exception is the low concentration, low luminosity case, where we may argue that the larger
fraction of central galaxies in the control sample causes their bluer colours.}
The mean and its associated error for each
of the presented distributions of galaxy colours (circles with error-bars and
Table 1) confirm 
all the trends mentioned above with high statistical
confidence levels.

{  The average values of stellar mass are comparable between normal, control and AGN host
galaxies for samples of similar luminosities and concentrations (see columns 7 and 8 in Table 1).
And even though the fraction of central galaxies shows significant differences in a few cases,
these do not affect our conclusions.  The cases are, (i) for
low luminosities and low concentrations, normal galaxies show a slightly larger central galaxy
fraction but even though this may correlate with the bluer colours shown by these galaxies
(central galaxies in this range of luminosity tend to be bluer than equal luminosity satellites),
this can not produce such a large difference in colour. (ii) In the same subsamples, the fraction
of central control galaxies is higher than for the AGN hosts, but this would only indicate that
the colours of control galaxies with equal fractions of central galaxies would be more
similar to those of the AGN hosts.
{ (iii) The remaining case corresponds to high concentration, low luminosity objects, where
control galaxies show a larger fraction of centrals than AGN hosts; however, if
these fractions were forced to be similar, the colour difference would be even more significant.}
}

\subsection{Remotion of local galaxy density and local halo effects }

In this section we explore whether
{\it at fixed values of median local density} there are residual variations
of galaxy properties as a function of the distance to the closest galaxy cluster,
and if so, whether the constraining of the host group luminosity is able to remove this
residual dependence.  {  Several works (e.g. Wang et al. 2009) have reported a
dependence of the red galaxy fraction with the distance to clusters; however, it has
been reported that there is a tight correlation between the local environment and
the distance to clusters (e.g. Balogh et al. 2004).  Therefore, our aim is to find out
whether there are residual variations in the galaxy colours with the distance to clusters
once the local density is controlled; as shown by Gonz\'alez \& Padilla (2009) there
is a remaining variation in a sample of $\Lambda$CDM semi-analytic galaxies.}

\begin{figure*}
\begin{picture}(430,475)
\put(0,135){\psfig{file=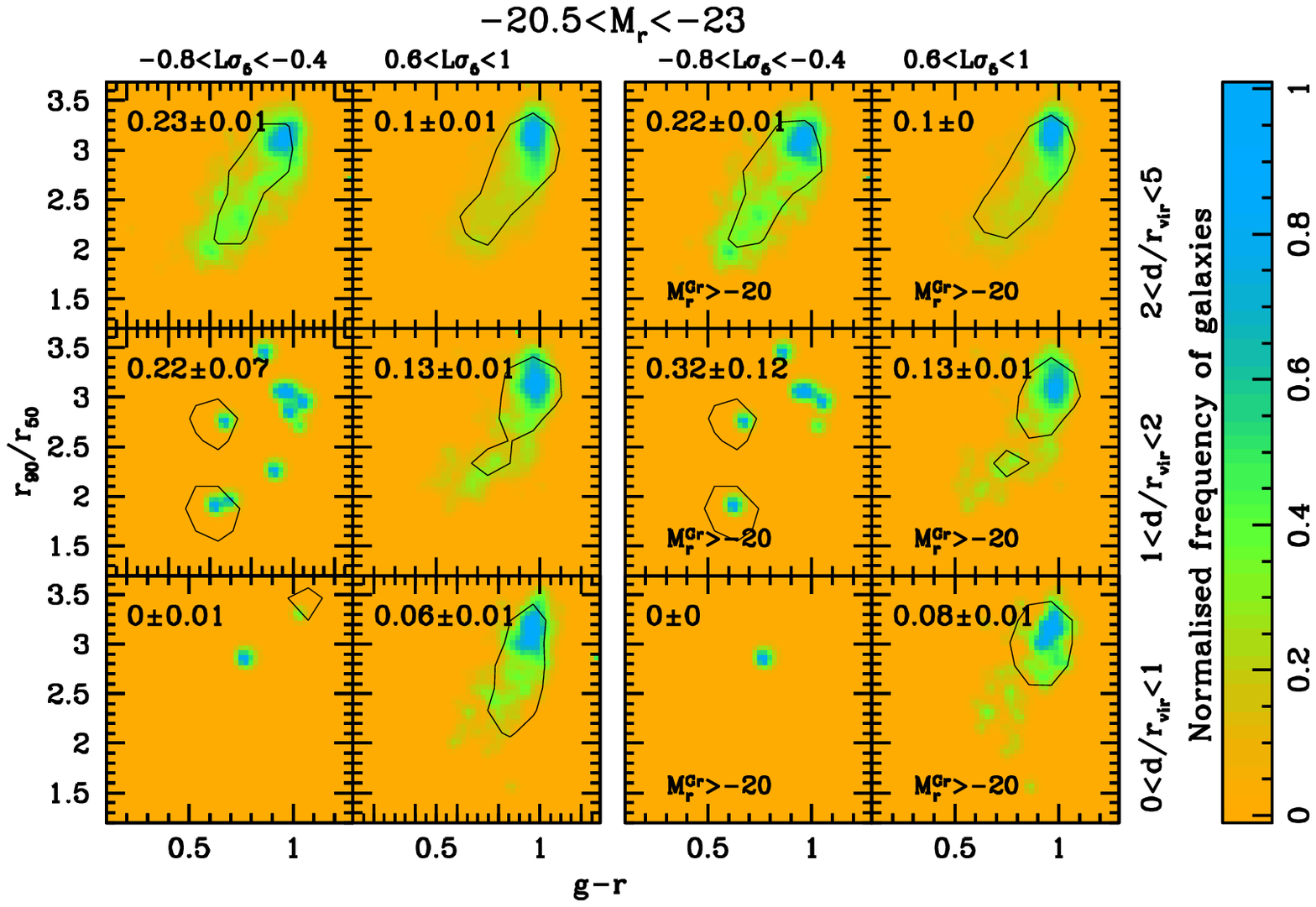,width=15.cm}}
\put(0,-110){\psfig{file=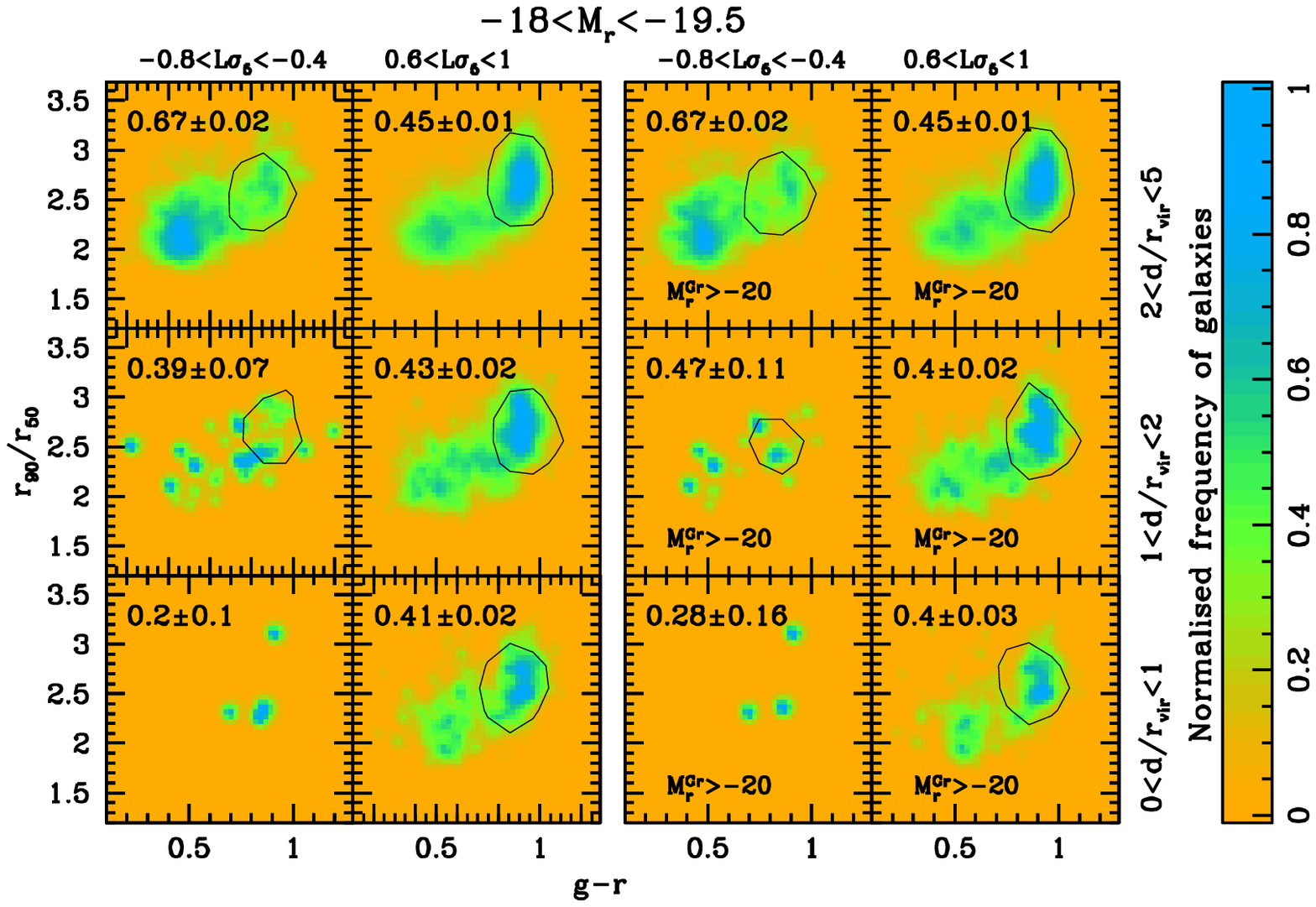,width=15.cm}}
\end{picture}
\caption{
Variations of galaxy properties as a function of distance to cluster centres for {\it fixed values of projected local
density}. Weighted distributions of galaxies (colours are as in Figure \ref{fig:paleta}) and AGN hosts (solid
lines containing $68$ percent of the total) for high, $-23<M_r<-20.5$, and low,
$-19.5<M_r<-18$, galaxy luminosities (top and bottom sets of panels, respectively).
Left and centre-left panels correspond to different ranges of local projected galaxy density 
(the legends on the top indicate the ranges). The centre-right and right panels  
correspond to the same local densities as in the two leftmost panels, with an additional
constrain on total host group luminosity of $M_r^{Gr}>-20$.
The numbers near the top
of each panel show the fraction of blue galaxies with $g-r<0.75$; the errors are Poisson, on the mean.
}
\label{fig:den}
\end{figure*}

{ Figure \ref{fig:den} shows concentration vs. colour diagrams 
in four panels divided in six sub-panels, each corresponding to different
combinations of parameters.  The top and bottom panels correspond to bright 
and faint galaxies ($-20.5<M_r<-23$ and $-18<M_r<-19.5$, respectively), and the
left and right panels correspond to unrestricted host group mass, and
host groups with $M_r^{Gr}>-20$, respectively.  In turn, each panel shows two
columns, the left for low local densities, $-0.8<\log_{10}(\sigma_5/$h$^{-2}$Mpc$^2)<-0.4$, 
and the right column for high local densities,  $0.6<\log_{10}(\sigma_5/$h$^{-2}$Mpc$^2)<1$;
the three rows in each panel correspond to different distances to the centres of clusters
(indicated to the right of the figure).
}

As can be seen in Figure \ref{fig:den}, 
when the local density is restricted to a narrow range, there is a small remnant of
the dependency reported from Figure \ref{fig:paleta}.$^($\footnote{
There is a small, negligible variation in the median local densities between the
samples shown, of $\Delta \log(\sigma_5/$h$/^{-2}$Mpc$^2)<0.05$, with the lowest 
median local density corresponding to the intermediate distance sample shown in Figure
\ref{fig:den}.}$^)$ Namely,
for low local densities, $-0.8<\log_{10}(\sigma_5/$h$^{-2}$Mpc$^2)<-0.4$, 
a small but significant blue population appears in the diagram for faint galaxies,
as the distance to the nearest cluster increases (left subpanels of bottom panel in Figure \ref{fig:den});
this is quantified by the fraction of blue galaxies shown in each subpanel, which increases
systematically with $d/r_{vir}$.
For the higher density samples, 
{ $0.6<\log_{10}(\sigma_5/$h$^{-2}$Mpc$^2)<1$,}
there are only small variations in the
bulge concentration vs. colour diagrams (second column of panels from the left); 
for faint galaxies (bottom left panels) 
the same effect is present with a higher statistical significance 
as indicated by the systematically increasing relative fraction of the blue, low concentration
population with respect to that of red galaxies as the distance to clusters increases
(indicated in each panel).
The fact that a dependency on cluster-centric distance is still marginally visible in the
samples with fixed local density indicates that the morphology-density relation found in previous
works such as Balogh et al. (2004), Baldry et al. (2006), O'Mill et al. (2008),
are in principle detecting only part of the possible dynamical range of variations
in the galaxy population.  
A more fundamental parametrisation should at least involve
the global environment of the galaxy as well as the local density.  Here we parameterised the
former by the distance to clusters of galaxies.  
This result is in agreement with the study of the residual
variations in the fraction of blue galaxies at fixed local densities but at increasingly
larger distances from cluster centres found by Gonz\'alez \& Padilla (2009)
in semi-analytic models.  It is also in agreement with studies of the colours
of satellite galaxies around groups in the SDSS by van den Bosch et al. (2008), where at
fixed stellar mass, satellites tend to be redder than central galaxies; this could be
the reason behind the decrease in blue fractions as the distance to clusters decreases modulo
local density effects.

The option of additionally fixing the mass of the DM host halo could provide the alternative
parametrisation needed to remove the dependence on cluster-centric distance.  
This was pointed out by Gonz\'alez \& Padilla (2009), who find that the
local and global density dependencies of the galaxy population can be seen in the distributions
of galaxy colours, but also as
variations in the mass of their DM halo hosts.  They also find a smaller residual
effect coming from the assembly of groups that we will not consider in the present analysis.
Gonz\'alez \& Padilla estimate that to 
some degree, this is in accordance with new results where it is found that the
evolution of galaxies takes place while they are part of groups, which later merge to
form larger structures such as clusters of galaxies (McGee et al. 2009; Wilman et al. 2008).

\begin{figure*}
\begin{picture}(340,360)
\put(-80,-45){\psfig{file=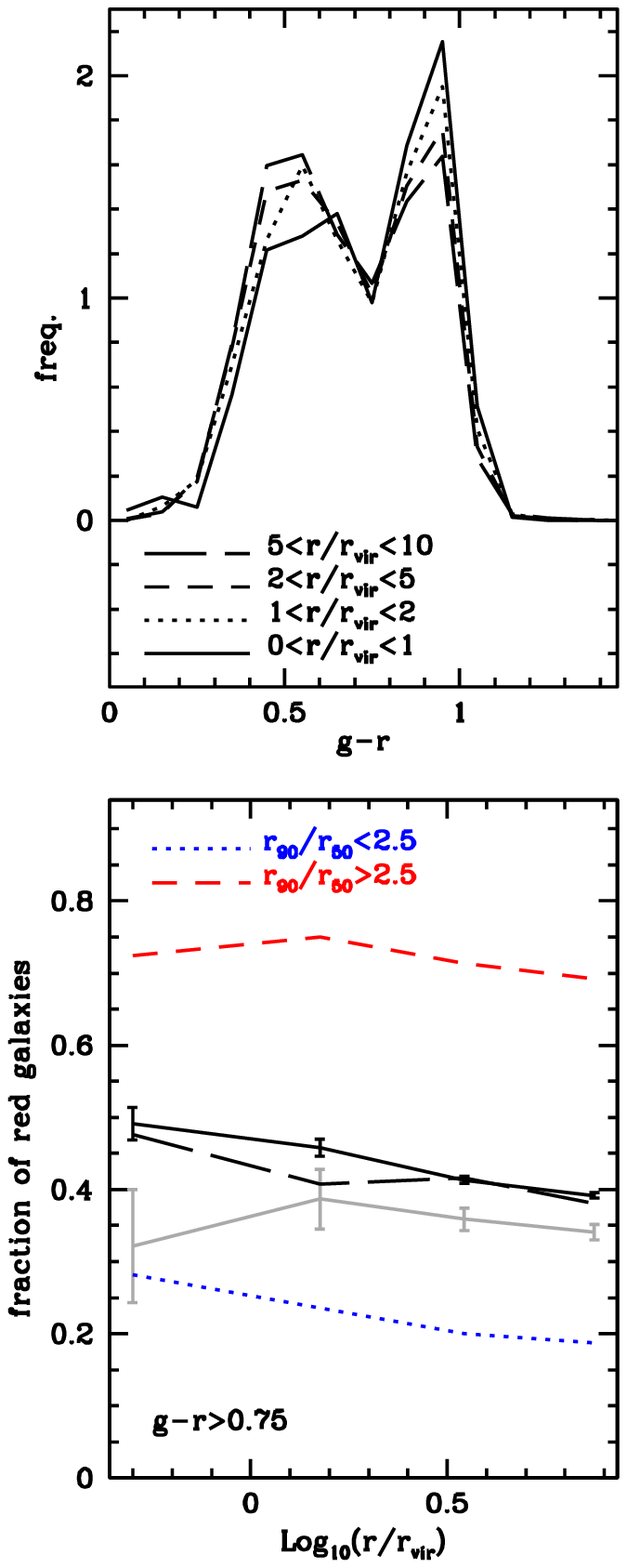,width=15.cm}}
\put(-40,-45){\psfig{file=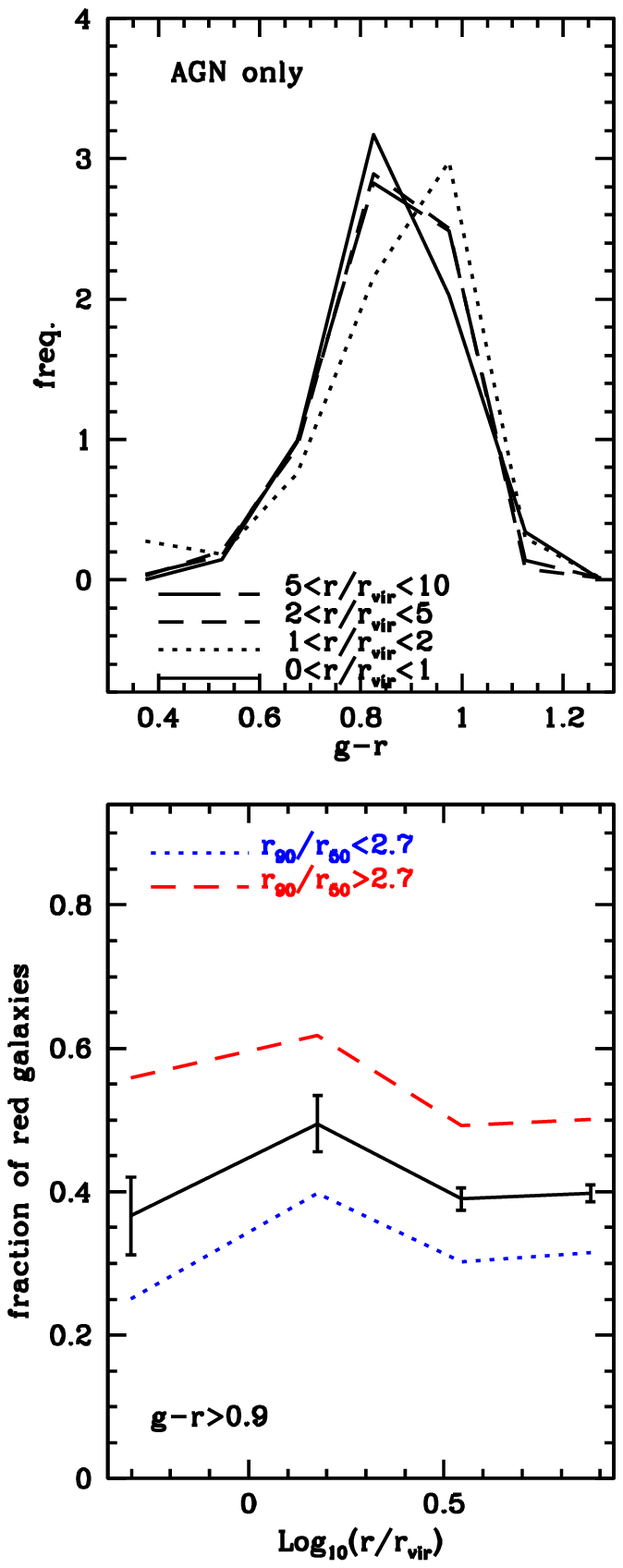,width=15.cm}}
\put(120,-45){\psfig{file=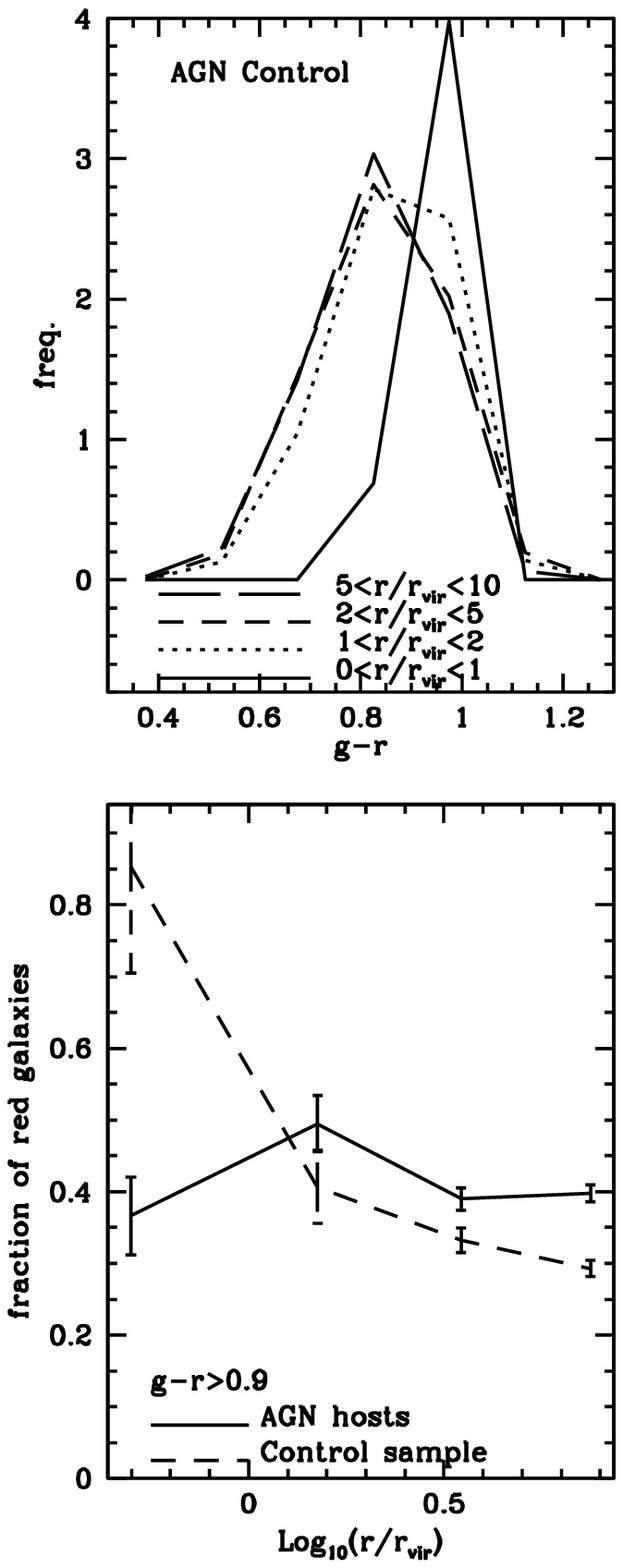,width=15.cm}}
\end{picture}
\caption{
Top panels: $1/V_{MAX}$ weighted distributions of colour
for galaxies in {\it equal local density environments}, but 
different cluster-centric distances (shown in different
line-types indicated in the figure key).  
The left panels are for non-AGN galaxies; centre panels correspond to the full AGN population;
right panels to the AGN control samples.
In all cases the local density is fixed at equal average values
by limiting $-0.3<\log(\sigma_5/h^{-2}Mpc^2)<0.3$.  
Lower-left and lower-centre panels: fraction of red galaxies as a function of
distance to the closest cluster of galaxies (solid lines)
regardless of concentration and for high (dashed) and low (dotted) concentrations.  The
cuts in colour and concentration are shown in the key.
The additional grey solid line in the lower left panel shows the fraction of red galaxies
as a function of distance to cluster centres for galaxies with an additional cut in
host group luminosity of $M_r^{Gr}>-20$.
The lower-right panel shows the fraction of red galaxies for the control sample (dashed)
and for the AGN hosts (solid, repeated from the lower-centre panel).  Errorbars correspond
to Poisson uncertainties on the mean.
}
\label{fig:globalrot}
\end{figure*}

We adopt our proxy for host halo mass for the individual galaxies described in
Section \ref{sec:data},
which consists of an assignment of the added luminosity of the four brightest group members.
We use the group luminosity as
previous findings point out that the group luminosity is a better proxy for the DM mass of a group
than its virial estimate (e.g. Eke et al. 2004).
The two right-most columns of sub-panels in Figure \ref{fig:den} show the resulting
galaxy populations in the bulge concentration vs. colour plane, for the same local density
limits used in the left panels, but for groups with faint luminosities ($M_r^{Gr}>-20$).
{ The measured median group luminosities are indistinguishable for both the low and high local density
samples and at different distances from clusters}. The results are shown for 
two different ranges of galaxy luminosity (top and bottom panels).  As can be
seen, { for low group luminosities (right panels),} 
it is slightly less clear  (as evidenced by analysing the
fractions of blue galaxies and their errors, indicated in the top left of each subpanel) that 
there are systematic shifts in the populations of galaxies 
as the cluster-centric distances increases { regardless of local density and galaxy luminosity}
(the only exception is that of faint
galaxies in low density environments).  
But due to the larger errors involved in this
analysis, it is only marginally preferred, at best, to use the host
DM halo mass in addition to the local density
as the ideal double parameter to separate different galaxy populations.
Larger data sets will be needed to rule out a possible dependence
on assembly.  

Finally, the AGN population does not show any obvious trends with
cluster-centric distance
when the local density or the host halo mass are constrained.  However, this may be
the result of low number statistics.  

Figure \ref{fig:globalrot} is shown to help quantify these points; in all the panels,
the local density is held fixed at $\sigma_5=1/h^{-2}Mpc^2$ (median value).  { 
The left panels correspond to the full galaxy population, the middle panels to
AGN hosts, and the right panels to the control sample.  The top panels show distributions
of $g-r$ colour, and the bottom panels fractions of red galaxies where the dividing colour
is shown on the bottom-left corner of each bottom panel.}

As in Figure
\ref{fig:rot}, the top-left panel
shows the $g-r$ colour distribution for galaxies with $-23<M_r<-18$ but this time
corresponding to
different distances from clusters (see the figure key), and only for galaxies
in the fixed moderate density environment to avoid
variations of galaxy properties coming from their local density.
The distributions show only a moderate bimodality with peaks located at roughly fixed positions,
with the red and blue peaks separated by 
$\Delta(g-r)=0.43\pm0.04$ as found in the analysis of samples in different local densities.
There is a hint of a {  shift towards higher red galaxy fractions} as the cluster-centric distance
diminishes.  This trend is quantified in the lower left panel, where solid lines show
the fraction of red ($g-r>0.75$) galaxies as a function of $r/r_{vir}$ (errorbars show Poisson
errors only for the sample with no restriction
on concentration), where this fraction drops by $0.099\pm0.011$ (Poisson uncertainty) 
between $0$ and $10$ virial radii
from the clusters of galaxies.  We also show this fraction in dotted lines for
low concentration galaxies, and in dashed lines for high concentration galaxies; we again find 
bluer colours of lower concentration galaxies.
{  Given that it has been pointed out that the host-halo mass may be a better indicator
of the local environment than $\sigma_5$ (e.g. Weinmann et al. 2006, in observations; Gonz\'alez
\& Padilla 2009, in simulations), we have also measured the fractions of red galaxies
as a function of distance to cluster centres for {\it fixed values of the host halo mass} 
(average total group luminosity of $M_r^{Gr}=-20.0$).  The result is shown as a black long-dashed
line in the lower left panel of the figure, and as can be seen we recover the results obtained
by fixing the local density (a variation in the red fraction of $0.096\pm0.019$).}

We make an additional subsample by requiring {  fixed local densities $\sigma_5=1/h^{-2}Mpc^2$ and }
low host group luminosities, $M_r^{Gr}>-20$,
and show the dependence of the red fractions with the distance to clusters as a grey solid line.
In this case the dependence is almost null, although the errorbars are considerably larger than
for the black solid line (for the grey line the local density is also held fixed).  Therefore the 
{  simultaneous constraint on the local density and the group luminosity} 
(proxy for mass) is able to marginally improve the selection of
a galaxy population fully independent of global environment, a result that is definitely not achieved
by fixing the local galaxy density {  or the host halo mass} alone.
Due to low number statistics we do not place further limits
on the masses of the host groups for AGNs or in the AGN control sample.

The top-centre panel of the figure shows the $g-r$ colour distributions of AGN hosts, 
for the full range of
nuclear OIII luminosity but with the same restriction on local density.  
As can be seen, both in the distributions and the red
fractions (shown in the lower-centre panel), AGN hosts are consistent
with no change in their colours as a function of distance to clusters once the local density is held
fixed,
at least to the degree of statistical certainty allowed by these observational data.
An interesting feature in the fraction of
red galaxies as a function of the distance to cluster centres is the bump at $r/r_{vir}\sim2$.
This would coincide with results from Coldwell et al. (2009) who also find an excess of AGN with 
red hosts in the outskirts of clusters.  However, the significance of this feature in our
analysis is only marginal, and it does not improve by changing the range
of masses of the cluster sample (we varied the low mass limit from $10^{13}$ to 
$5\times10^{14}h^{-1}M_{\odot}$).  

Galaxies in the control sample show a clear dependence of colours as a function of 
the distance to
clusters (at fixed local density) as can be seen in the right panels (a total variation
of $0.56\pm0.07$ in the fraction of red control galaxies).  The lower-right
panel shows in a dashed line the dependence of the red galaxy fraction on distance to clusters for
the control sample, which can be compared to the dependence obtained for the AGN sample (shown
in this panel as a solid line,
repeated from the lower-centre panel).  It is clear that the
colours of AGN control galaxies (at a fixed local density, $\sigma_5$) 
show a clear variation towards bluer colours as the distance to the centres of clusters of galaxies
increases.
{ Even though AGN show almost constant red fractions as a function of distance to cluster centres,
relative to control galaxies, AGN hosts are significantly redder}
for $r/r_{vir}>2$, where the fraction of red galaxies is $0.313\pm0.015$ and
$0.394\pm0.014$ for the control and AGN samples, respectively
and significantly bluer for $r/r_{vir}<2$, with fractions of red galaxies of $0.63\pm0.11$ and
$0.39\pm0.14$, respectively
 with comparable stellar masses and
fractions of central galaxies.  In particular, $100\%$ and $93\%$ of AGN hosts and control
galaxies at $r/r_{vir}<1$ are satellites ($72\%$ and $80\%$, respectively, 
at $r/r_{vir}>5$) in this figure. 
This could indicate that towards the centres of clusters, {  satellite} galaxies 
{  at fixed local density environments} need to be more capable of feeding
a central black hole to be detected as AGN, than outside clusters { (consistent
with the constant red fractions as a function to cluster centres shown by AGN hosts)};
this would point towards the known SF-AGN activity relation (Kauffmann et al. 2004) {  but
while the properties of the galaxy hosts are the same, including their environment}.  
The result by von der Linden et al. (2009) who
obtain a reddening of the hosts of AGN towards the cluster centres, could be detecting
the variation of the properties of typical AGN hosts as measured by our control samples.

We remind the reader
that these latter results correspond to an environment characterised by a fixed median local density
of $\sigma_5=1/h^{-2}Mpc^2$.  { When analysing higher densities, $\sigma_5>10/h^{-2}Mpc^2$,
we find consistent results, but with AGN colours bluer than galaxies in the control sample out to
greater distances from the cluster centres (probably due to the effect of higher local densities);
in this case, the group luminosities that produce similar fractions of red galaxies than this
higher local density cut are $-24<M^{Gr}_r<-22$.  Our
data set does not allow us to draw conclusions for low densities, $\sigma_5<0.5/h^{-2}Mpc^2$, since
the resulting samples contain too few galaxies. }

\section{Discussion}

{ 
The aims of this paper consisted on studying the variations of the properties
of galaxies with their environment, and to compare it to what is expected from
galaxy formation models.  We have analysed separately these variations for normal
galaxies and for galaxies with AGN.  We will now compare our results
to previous studies for these two types of galaxies in the following subsections.

\subsection{Results for the full galaxy population}

There have been numerous studies on the environmental dependence of galaxy properties,
where the environment has been given for example by projected galaxy densities on different
scales (e.g. Balogh et al. 2004; Baldry et al. 2005; Kauffmann et al. 2004; O'Mill, 
Padilla \& Lambas 2006), the host halo mass (Weinmann et al. 2006), or 
the distance to cluster centres (e.g. Wang et al. 2009; Smriti \& Raychaudhury 2009).  As these three
quantities are correlated, the results obtained by different authors consistently show
that galaxies located in lower density environments, or
further away from clusters, or in host haloes with lower masses, are characterised by
bluer colours.  

Given that it is expected that 
the local density
in the very high redshift Universe 
affects the final local dark-matter halo mass (e.g. Weinmann et al. 2006; Gonz\'alez
\& Padilla 2009), and in turn, the history of mergers, it could be expected
that the properties of galaxies at {\it equal $z=0$ local densities} but different
large-scale environments tracing different initial conditions, could show different
properties.  Note that this would also be the case for galaxies with {\it equal
host-halo masses}, as the different merger histories, could also imprint different
galaxy properties in haloes of equal mass, but different large-scale environments { and 
therefore, different assembly histories}.
This is clearly expected in view of the assembly bias effect, now detected
in simulations (starting with Gao, Springel \& White 2005) and observations
(Wang et al. 2008; Zapata et al. 2009), where equal mass haloes with different
assembly histories show different clustering properties as well as different
galaxy populations.

Therefore, we studied the variation of red galaxy fractions as the distance
to clusters increases, but at {\it fixed values of local densities or host-halo masses},
and found an increasing fraction of blue galaxies with a high statistical significance
($9\sigma$ and $5\sigma$ for fixed local densities and host-halo masses, respectively).  
With this result, we confirmed
those found for semi-analytic galaxies by Gonz\'alez \& Padilla (2009).  Previously,
Ceccarelli, Padilla \& Lambas (2008) had been able to find an indication of this
effect, but with much lower statistical significance, comparing galaxies in equal
local density environments, but in void walls, and outside voids.

Gonz\'alez \& Padilla (2009) also found that even when the local densities and { 
DM halo} masses are constrained
to fixed values, different large-scale environments can still induce changes on the
modelled galaxy properties, an effect that could be attributed to  coherent motions
or further assembly effects.  We repeated this analysis with our full galaxy sample,
but were not able to find any significant remaining variations of colours.  Given the large
errorbars involved, larger galaxy samples are needed in order to confirm this effect in
observations.

\subsection{Results on AGN hosts}

Our analysis of the AGN hosts used projected local densities, host halo masses and
their distance to clusters of galaxies (and combinations of these) as
proxies for their environment.  We also defined a control sample of non-AGNs such that their
stellar masses, r-band luminosities, colours, local densities, concentrations, fractions
of central and satellite galaxies, and host halo masses, show the same distributions as
the AGN hosts.  By applying all these restrictions
independently on each parameter of non-AGN galaxies,
any correlations between these quantities introduced by, or allowing the, AGN activity
will produce differences with the control sample, particularly when analysing different environments.
In the literature, several works make comparisons between the properties of AGN hosts and of non-AGN galaxies
characterised by similar properties, but still allowing for differences on several parameters.
For instance, most 
studies on the fractions of AGN 
essentially make a comparison to the full galaxy population
divided according to local density, host morphology, stellar mass, and nuclear OIII luminosity
(Miller et al. 2003; Kauffmann et al. 2004;
Popesso \& Biviano 2006; Cappi et al. 2001; Molnar et al. 2002; Martini et al. 2002;
Martini, Sivakoff \& Mulchaey 2009; Pasquali, van den Bosch \& Rix 2007).
In other cases, the studies served to identify the environments of AGN hosts
(Choi et al. 2009; Waskett et al. 2005; Gilmour et al. 2007; Silverman et al. 2009; Martel
et al. 2007; von der Linden et al. 2009; Montero-Dorta et al. 2009; Koulouridis et al. 2006;
Martini et al. 2004) using a single parameter such as local density, distance to cluster centres
or cluster membership;
or the typical AGN host properties (Kauffmann et al. 2003a; Kewley et al. 2006; Hao et al. 2009),
as in Pasquali et al. (2009) who studied the hosts of AGN as a function of the host halo mass.
In the process of constructing our control samples, we compared the distributions of several
parameters of AGN hosts and the full galaxy population, and we were able to confirm several
of these results (for example, constant fraction of AGN as a function of local density, a larger
occurrence of AGN in higher mass host haloes).

Our approach is complementary to these previous works, and is designed to find the effect
of changing only one parameter, the detection of AGN activity.
In our study we found that as the local density or the host halo mass increase, AGN hosts become
slightly redder (while occupying the green valley) and of higher luminosity, consistent
with studies of the hosts of AGN.  However, even though the control galaxies
also show this qualitative behaviour, they are always slightly bluer and fainter than AGN hosts,
a clear difference coming from a condition related to the AGN detection, either an effect
of the BH on the galaxy, or a condition that allows the BH activity.  In this sense,
this could indicate that AGN in clusters are hosted by larger galaxies (a correlation in
the AGN hosts not present in the control samples) in order to be able to retain gas in
the intracluster conditions. 
Note that this result is seemingly
contrary to what was found by Pasquali, Van den Bosch \& Rix (2007, { although
this paper deals only with early type galaxies}) and Hao et al. (2007), since
they connect the AGN activity to star formation events (disky and barred galaxies, respectively).
However, part of this signal is due to the typical hosts of AGN, and not to the detection of AGN
activity in these galaxies.

Consistent
with the interpretation that our control samples contain dormant black holes where the SF
has regained its strength, we find that when the local density
is held fixed at intermediate (or high) values and the distance to clusters varies, the hosts
of AGN { show constant colours.  Thus, close to the clusters AGN are significantly bluer 
than the control galaxies } (however, it could
also be possible that AGN near cluster centres need more available gas to remain active).
In this case most of the galaxies (AGN hosts and control) are satellites of the cluster and
therefore the AGN hosts under analysis would only conform to a subsample of the AGN
studied by Pasquali et al. (2009), who do not divide their sample according to local density.

At low local densities, 
the low OIII luminosity AGN are found to extend their range of colours and concentration
to lower values as the host luminosity increases.  Only part of this behaviour comes from
the morphologies and characteristics of the AGN hosts, as the control sample shows a less
clear trend, in slightly better agreement with the behaviour of normal galaxies (higher
concentrations and colours).  This could favour the view that the AGN are not the
most important factor in transforming galaxies into red and dead objects as suggested
by Schawinski et al. (2007), in comparison to cluster environmental processes (Kewley et al. 2006)
since at low densities low OIII luminosity AGNs can even show blue colours.

\subsection{Perspectives}

Even though we were able to detect signatures of assembly on the properties of galaxies
for our full sample, in the form of different colours for galaxies at fixed local densities and
different distances from clusters, there still remain many questions regarding the effect from
particular details of the development of structures in the Universe.  
For instance, the apparent correlation between
coherent motions and the remnant variations of galaxy properties at fixed local density
and host halo mass (Gonz\'alez \& Padilla 2009), or the contamination of the areas surrounding
clusters by ejected satellite galaxies (Wang et al. 2009).  This will require considerably
larger galaxy samples, which could also be useful in detecting the evolution of this
phenomena with redshift.   These effects will be small but they are
an expected outcome of the development of galaxies within a $\Lambda$CDM scenario. 

The same applies to the study of AGN hosts, where the comparison to a strict control sample
has allowed us to find the correlations between the detection of an AGN and the local density,
the distance to clusters (with respect to a control sample), and the host 
halo mass.  More systematic studies
on the effect of each parameter included in the selection of the control sample are possible
and would allow a more detailed understanding of the processes behind the detection of an AGN
and its relation to the different phenomena associated to the origin and development of structure.
}

\section*{Acknowledgments}
We acknowledge constructive comments from Franz Bauer and an anonymous Referee.
This work was supported in part by the FONDAP ``Centro de Astrof\'\i sica" $15010003$, BASAL-CATA,
Consejo Nacional de Ciencia y Tecnolog\'\i a (PIP 6446), and Agencia Nacional de 
Promoci\'on Cient\'\i fica y 
T\'ecnica (PICT 32342).  NP was supported by a Proyecto Fondecyt Regular No. 1071006.
DGL acknowledges travel support to Santiago de Chile 
from Proyectos Fondecyt Internacional No. 7070044 and 7080131.
RG acknowledges receipt of a MECESUP PUC0609 Fellowship and support from 
Fondo ALMA-CONICYT 31070007.

\bsp

\label{lastpage}

\end{document}